\documentclass[reqno,11pt]{amsart}
\usepackage{graphicx}
\usepackage{slashed}
\usepackage{amscd}
\usepackage{tensor}
\usepackage{amsmath,amssymb,mathabx,bm}
\usepackage{esint}
\usepackage{dsfont}
\usepackage{enumitem}
\usepackage{accents}
\usepackage[usenames,dvipsnames]{pstricks}
\usepackage{epsfig}
\usepackage{pst-grad} 
\usepackage{pst-plot} 
\usepackage[mathscr]{eucal}
\numberwithin{equation}{section}
\allowdisplaybreaks[1]

\title[Causal Fermion Systems and the ETH Approach]{Causal Fermion Systems and the \\
ETH Approach to Quantum Theory}

\author[F.\ Finster]{Felix Finster}
\address{Fakult\"at f\"ur Mathematik \\ Universit\"at Regensburg \\ D-93040 Regensburg \\ Germany}
\email{finster@ur.de}
\author[J.\ Fr\"ohlich]{J\"urg Fr\"ohlich}
\address{Institute of Theoretical Physics, ETH Zurich, Switzerland}
\email{juerg@phys.ethz.ch}
\author[M.\ Oppio]{Marco Oppio}
\address{Fakult\"at f\"ur Mathematik \\ Universit\"at Regensburg \\ D-93040 Regensburg \\ Germany}
\email{marco.oppio@ur.de}
\author[C.F.\ Paganini]{Claudio F. Paganini \\ \\ April 2020}
\address{Fakult\"at f\"ur Mathematik \\ Universit\"at Regensburg \\ D-93040 Regensburg \\ Germany}
\address{Max Planck Institute for Gravitational Physics (Albert Einstein Institute), Am M\"uh\-len\-berg 1, D-14476 Potsdam, Germany}
\email{claudio.paganini@ur.com}

\newtheorem{Def}{Definition}[section]
\newtheorem{Thm}[Def]{Theorem}

\newtheorem{axiom}{Axiom}
\newcommand{\boundedlin}{\Lin}
\newcommand{\Thanks}{\vspace*{.5em} \noindent \thanks}
\newcommand{\beq}{\begin{equation}}
\newcommand{\eeq}{\end{equation}}
\newcommand{\Proof}{\begin{proof}}
\newcommand{\QED}{\end{proof} \noindent}

\newcommand{\la}{\langle}
\newcommand{\ra}{\rangle}

\newcommand{\C}{\mathbb{C}}
\newcommand{\R}{\mathbb{R}}

\newcommand{\N}{\mathbb{N}}

\DeclareMathOperator{\tr}{tr}
\renewcommand{\O}{{\mathscr{O}}}
\renewcommand{\L}{{\mathcal{L}}}
\newcommand{\Sact}{{\mathcal{S}}}

\DeclareMathOperator{\supp}{supp}
\renewcommand{\H}{\mathscr{H}}

\newcommand{\Lin}{\text{\rm{L}}}
\newcommand{\F}{{\mathscr{F}}}

\newcommand{\A}{\mycal A}

\newcommand{\bitem}{\begin{itemize}[leftmargin=2.5em]}
\newcommand{\eitem}{\end{itemize}}

\DeclareFontFamily{OT1}{rsfso}{}
\DeclareFontShape{OT1}{rsfso}{m}{n}{ <-7> rsfso5 <7-10> rsfso7 <10-> rsfso10}{}
\DeclareMathAlphabet{\mycal}{OT1}{rsfso}{m}{n}

\begin{document}

\maketitle

\begin{abstract} 
After reviewing the theory of ``causal fermion systems'' (CFS theory) and the ``Events, Trees, and Histories Approach'' to
quantum theory (ETH approach), we compare some of the mathematical structures underlying these two general frameworks and discuss similarities and differences.
For causal fermion systems, we introduce future algebras based on causal relations inherent to a
causal fermion system. These algebras are
analogous to the algebras previously introduced in the ETH approach.
We then show that the spacetime points of a causal fermion system have properties similar to those of ``events'', 
as defined in the ETH approach.
Our discussion is underpinned by a survey of results on causal fermion systems describing Minkowski space that
show that an operator representing a spacetime point commutes with the algebra in its causal future, up to tiny corrections
that depend on a regularization length.
\end{abstract}

\tableofcontents

\newpage
\section{Introduction}\label{sec:introduction}
Two fundamental problems in theoretical physics are (1) to find a formulation of quantum theory (QT) incorporating a precise notion of \textit{events} that leads to a solution of the so-called \textit{measurement problem} and (2) to construct a theory unifying General Relativity (GR) and the Standard Model (SM) of particle physics. In this survey we sketch two different theoretical frameworks designed to address one of these two problems, each. 
Our goal is to work out the similarities and differences of the two approaches.
Indeed, when comparing their mathematical structures, we find rather striking analogies and similarities. This is somewhat surprising, because the two frameworks have been developed to solve different problems.

The two frameworks are the theory of \textit{``causal fermion systems''} (in short CFS theory) 
and the \textit{``Events, Trees, and Histories Approach''} to quantum theory (ETH approach).
The central postulate of CFS theory is the \textit{``causal action principle,''} from which the classical field equations of GR and of the SM can be derived in the continuum limit (see the textbook~\cite{cfs}, the reviews~\cite{dice2014, review}
or the website~\cite{cfsweblink}). The theory thus provides a uni\-fi\-cation of the weak, the strong and the electromagnetic forces with gravity
at the level of classical field theory. The approach also aims at incorporating full Quantum Theory. 
For the time being, it
has provided concise notions of ``quantum spacetime'' and ``quantum geometry''
(see~\cite{rrev, lqg}), and it has revealed connections to Quantum Field Theory (see~\cite{qft, fockbosonic}).
A basic idea underlying this approach is to describe all spacetime structures in terms of
linear operators on a Hilbert space.

The central postulate of the ETH approach (as initiated in~\cite{frohlich2015math, froehlich2019review, frohlich2016quest})
is the so-called \textit{``Principle of Diminishing Potentialities''} (PDP), which enables one to propose a precise 
definition of \textit{isolated open systems}, come up with a compelling notion of  {\em{events}} featured by such systems, 
and to understand the nature of the stochastic \textit{time evolution of states} in quantum theory. Among central 
ingredients of the ETH approach are algebras, $\mathcal{E}_{\geq t}$, generated by bounded operators 
representing potential events that may happen at or after some future time $t$. 
The ETH approach is intended to represent a completion of QT enabling one to solve fundamental conceptual problems, such as the infamous \textit{measurement problem}.

The two frameworks, CFS theory and the ETH approach, show striking similarities at the conceptual level and in 
their mathematical structure.
A basic idea inherent in the ETH approach, but not fully understood at present, is that spacetime 
is \textit{woven by events}, in a sense first suggested by Leibniz: Spacetime and its causal structure encode the location of \textit{events} and \textit{relations among events}. In a relativistic formulation of the ETH approach, an event is represented by an orthogonal projection localized in a spacetime cell (see Definition~\ref{ETHevent}). 
Similarly, in CFS theory, spacetime structures are described in terms of ``relations among operators.''
More precisely, each spacetime point is described by an operator on a Hilbert space,
and spacetime as well as all structures therein (like the geometry and matter fields)
are derived from these operators by using, for example, spectral 
properties of products of such operators (see Section~\ref{seccfsgen}).
Among similarities between the mathematical structures underlying CFS theory and the ETH approach, 
we note that operator algebras play somewhat analogous roles in the two frameworks
and that the operators describing spacetime points of a causal fermion system are reminiscent of the 
operators representing events in the ETH approach.

Our goal in this survey is to identify and discuss similarities and differences between the two approaches.
We start our analysis by presenting brief, but concise introductions to the two frameworks, which serve a two-fold 
purpose. First, they are supposed to make this paper understandable to readers who are not familiar with either 
framework; and, second, they facilitate a comparison between the two frameworks, the ETH approach and CFS theory.
A rather substantial effort is devoted to a detailed comparison of the mathematical structures underlying the two 
frameworks and of the way dynamics is treated. This can be considered to be the central purpose of our survey. A discussion of some recent mathematical results for causal fermion systems in Minkowski space complements the comparison of CFS theory with the ETH approach. 

We emphasize up front that we have not succeeded and do not intend to merge CFS theory with the ETH approach. However, in Section 6, the prospects for success of a research program are discussed that attempts to embed both these frameworks in a common one that encompasses them both. We hope that this survey will stimulate further research in this direction.

\subsubsection*{Organization of the paper}
In the Section~\ref{sec:CFS}, we outline the theory of causal fermion systems. 
In Section \ref{sec:ETH}, we present an introduction to the ETH approach to QT. In Section \ref{sec:comparison}, we attempt to compare, step by step, the structures underlying CFS theory and the ETH approach to QT. In Section \ref{sec:results}, we sketch some recent results in the analysis of causal fermion systems in Minkowski space. 
In Section \ref{sec:outlook},  we conclude our analysis with some speculations about how incompatibilities between the 
present formulations of CFS theory and the ETH approach to QT might be eliminated, and in which direction one might search for a larger theoretical framework unifying the two theories.

\section{Basics on Causal Fermion Systems}\label{sec:CFS}
\subsection{General Setup} \label{seccfsgen}
Assuming that many readers are not familiar with causal fermion systems,
we give a brief overview of the basic concepts of the theory and explain how they relate to 
physical concepts familiar to most readers.
For a more detailed introduction tailored to the taste of physicists, see~\cite{dice2014,dice2018}.
Reference ~\cite{cfsrev} might be a good starting point for mathematicians.
A comprehensive treatment of the theory is given in the textbook~\cite{cfs}.

A basic goal at the root of the theory of causal fermion systems is to encode spacetime, 
its entire geometrical structure as well as matter
in a \textit{measure} defined on a family of symmetric operators acting 
on a Hilbert space. We start our discussion by presenting the abstract definition of a causal fermion system.

The algebra of bounded linear operators on a Hilbert space~$\H$ is denoted by~$\Lin(\H)$.
\begin{Def} \label{def:cfs}
Let~${(\H, \la .|. \ra_\H)}$ be a Hilbert space. 
Given a natural number {$n \in \N$} (a parameter called ``{spin dimension}''), we define a set $\F \subset \Lin(\H)$ by 
\begin{align*}
{\F} := \Big\{ \;x &\in \Lin(\H)\, \vert\, x \text{ has the properties:} \\
&\blacktriangleright\; \text{$x$ is {selfadjoint} and has {finite rank}} \\
&\blacktriangleright\; \text{$x$ has at most $n$ positive and at most $n$ negative eigenvalues}\,\Big\}  \:.
\end{align*}
Let~$\rho$ be a measure on~$\F$.
We refer to~$(\H, \F, \rho)$ as a {\bf{causal fermion system}}.
\end{Def} \noindent
The connection of this definition to physics is not at all obvious. For this reason, we explain its basic ingredients and their physical interpretation. 
In Section~\ref{sec:continuum} the connection of causal fermion systems to physics is explained for 
the example of Minkowski space.

The measure~$\rho$ associates to subsets $\Omega \subset \F$ a non-negative number, their ``volume''~$\rho(\Omega)$.
The introduction of this measure has two purposes.
\bitem
\item[(i)]{It gives rise to a notion of spacetime. To this end,
we introduce the notion of  support\footnote{The
{\em{support}} of a measure is defined as the complement of the largest open set of measure zero, i.e.,
\[ \supp \rho := \F \setminus \bigcup \big\{ \text{$\Omega \subset \F$ \,\big|\,
$\Omega$ is open and $\rho(\Omega)=0$} \big\} \:. \]}, of $\rho$, 
\[  M := \supp \rho \:. \]
Points of~$M$ are called {\em{spacetime points}}, and~$M$ is referred to as the {\em{spacetime}} of the causal fermion system~$(\H, \F, \rho)$.}
\item[(ii)]{Second, $\rho$ defines a volume element on spacetime: for ~$\Omega \subset M$,
~$\rho(\Omega)$ can be thought of as the volume of $\Omega$, thus generalizing the usual four-dimensional volume of a subset in spacetime.}
\eitem
The notion of spacetime introduced here turns out to be quite different from standard Lorentzian spacetime. 
Many familiar features, such as the one of a smooth structure, metric, etc.\ are absent in the notion of spacetime 
used in the theory of causal fermion systems.
Spacetime points $x, y, \dots$ in $M \subset \Lin(\H)$ of a causal fermion system are \textit{linear operators} 
on $\H$, and hence they encode much \textit{ richer} information, namely the eigenvalues and eigenvectors of $x$, than a traditional spacetime point. This additional information is supposed to encapsulate the entire structure of spacetime \textit{and} matter.
In order to convince readers that the notion of spacetime and spacetime points used in the theory of causal fermion systems is sensible, one should show how, starting
from the abstract definition of a causal fermion system, one can recover, in suitable limiting regimes, the usual structures defined on Lorentzian spacetime.
To this end, given a causal fermion system, one constructs various mathematical objects that are
inherent in the sense that they only require information encoded in the data, $(\H, \F, \rho)$, defining the
causal fermion system.
One then shows that, in some well-defined limiting regime, these objects correspond to ones that are familiar when one deals with a Lorentzian spacetime.
Such constructions are carried out in detail in~\cite{cfs}. They give rise to structures such as {\em{spin spaces}} and mathematical objects such as {\em{wave functions}},
{\em{connection}} and {\em{curvature}}. In what follows, we only discuss those objects and structures that are essential
for the analysis presented in this paper.

In order to define a {\em{causal structure}} on spacetime,
we study the spectral properties of the operator product $xy$, where $x$ and $y$ are spacetime points in $M \subset \F$. Note that this operator product is an operator
of rank at most~$2n$, where $n$ is the spin dimension of $\F$; in general it is not symmetric (because~$(xy)^* = yx \neq xy$ unless the
operators commute) and, thus, it is not an element of~$\F$.
We denote the non-trivial eigenvalues of~$xy$, counting algebraic multiplicities,
by~$\lambda^{xy}_1, \ldots, \lambda^{xy}_{2n} \in \C$. 
More precisely,
denoting the rank of~$xy$ by~$k \leq 2n$, we denote by~$\lambda^{xy}_1, \ldots, \lambda^{xy}_{k}$ all
the non-zero eigenvalues of $xy$ and set~$\lambda^{xy}_{k+1}, \ldots, \lambda^{xy}_{2n}=0$.
These eigenvalues give rise to the following notion of a causal structure.
\begin{Def} \label{def:causalstructure}
Two points~$x,y \in \F$ are said to be
\[ \left\{ \begin{array}{cll}
\text{{\bf{spacelike}} separated} &\!&  {\mbox{if $|\lambda^{xy}_j|=|\lambda^{xy}_k|$
for all~$j,k=1,\ldots, 2n$}} \\[0.4em]
\text{{\bf{timelike}} separated} &&{\mbox{if $\lambda^{xy}_1, \ldots, \lambda^{xy}_{2n}$ are all real}} \\[0.2em]
&& \text{and $|\lambda^{xy}_j| \neq |\lambda^{xy}_k|$ for some~$j,k$} \\[0.2em]
\text{{\bf{lightlike}} separated} && \text{otherwise}\:.
\end{array} \right. \]
\end{Def} 
\noindent More specifically, two points are lightlike separated if not all the eigenvalues have the same
absolute value and if 
not all of them are real. This notion of causality does not rely on an underlying metric. Yet, 
it reduces to the causal structure of Minkowski space or of curved spacetime in certain limiting
regimes (for Minkowski space, see~\cite[Section~1.2]{cfs} and Section~\ref{sec:continuum};
for curved spacetime, see~\cite{lqg}).

A causal fermion system introduces an {\em{arrow of time}}. It is convenient to use the notation
\begin{align}
S_x &:= \text{ range of } x = x(\H) \subset \H &&\hspace*{-1cm}\,\, \text{ a subspace of dimension $\leq 2n$} \label{Sxdef} \\
\pi_x &: \H \rightarrow S_x \subset \H &&\hspace*{-1cm} \,\,\text{ orthogonal projection onto } S_x \:.
\end{align}
The space~$S_x$ is referred to as the {\em{spin space}} at the point ~$x$.
This definition enables us to introduce the functional
\[ {\mathscr{C}} :  M \times M \rightarrow \R\:,\qquad
{\mathscr{C}}(x, y) := i \,\tr \big( y\,x \,\pi_y\, \pi_x - x\,y\,\pi_x \,\pi_y \big) \:, \]
which leads to the following notion of an arrow of time (for details see~\cite[Section~1.1.2]{cfs}).
\begin{Def}\label{def:order}
For timelike separated points~$x,y \in M$,
\[ \left\{ \begin{array}{cll}
\text{$y$ lies in the {\bf{future}} of~$x$} &\!&  {\mbox{if ${\mathscr{C}}(x,y)>0$}}\:, \\[0.4em]
\text{$y$ lies in the {\bf{past}} of~$x$} && {\mbox{if ${\mathscr{C}}(x,y)<0\:.$}}
\end{array} \right. \]
\end{Def} \noindent
We point out that this notion of ``lies in the future of'' is not necessarily transitive.
This is in accordance with the observation that the transitivity of causal relations
might be violated on the cosmological scale (there might exist closed timelike curves)
\textit{and} on the microscopic scale (there does not appear to be a compelling reason why causal
relations ought to be transitive at very tiny scales, down to the Planck scale).
We note, however, that causal fermion systems give rise to some other notions of causality
that \textit{are} transitive; we refer the reader to~\cite[Section~4]{linhyp}. 

After these preliminaries, we now come to the core of the theory of causal fermion systems, the {\bf{causal action principle}}.
In order to single out physically admissible causal fermion systems, one must formulate constraints 
in the form of physical equations. This is accomplished by postulating that
the measure~$\rho$ be a minimizer of the causal action,~${\mathcal{S}}$, which is defined as follows,
\begin{align}
\text{\em{Lagrangian:}} && \L(x,y) &= \frac{1}{4n} \sum_{i,j=1}^{2n} \big(
|\lambda^{xy}_i| - |\lambda^{xy}_j| \big)^2  \label{Lagrange} \\
\text{\em{causal action:}} && \Sact(\rho) &= \iint_{\F \times \F} \L(x,y)\: d\rho(x)\, d\rho(y) \:, \label{Sdef}
\end{align}
with some constraints to be added that are not made explicit here (these constraints are spelled out in detail
in~\cite[Sec\-tion~1.1.1]{cfs}). The right class in which to vary the measure~$\rho$
are the {\em{regular Borel measures}} on~$\F$, corresponding to the topology
induced by the operator norm on~$\Lin(\H)$.
A measure minimizing the action $\Sact$ satisfies Euler-Lagrange equations (see~\cite[Section~1.4.1]{cfs}).
These equations encode the dynamics of the causal fermion system.
In a suitable limiting regime, referred to as the {\em{continuum limit}}, this dynamics can
be expressed in terms of Dirac particles and anti-particles in Minkowski space interacting with classical fields, such as
the electromagnetic and/or gravitational fields; for details see~\cite{cfs}.

Before going on, we make a remark on the existence theory and explain the
connection to a class of variational principles referred to as {\em{causal variational principles}}.
Causal variational principles are a generalization of the causal action principle where
one replaces~$\F$ by a smooth manifold (compact or non-compact) and~$\L : \F \times \F \rightarrow \R^+_0$ by
a lower semi-continuous function (for details see~\cite{support, jet}).
If~$\F$ is compact, the existence of minimizers can be proved with the direct method
of the calculus of variations using the Banach-Alaoglu theorem (see~\cite[Section~1]{continuum} or~\cite{support}).
These methods also apply to the causal action principle in the case that~$\H$ is finite-dimensional
(see~\cite[Section~2]{continuum}).
If~$\H$ is infinite-dimensional, also the total volume~$\rho(\F)$ must be
infinite (see~\cite[Exercise~1.3]{cfs}). In this case, existence of minimizers has been proven
only under additional symmetry and compactness assumptions (see~\cite[Section~4]{continuum}),
but the general case is still open.
A first step in this direction is the existence theory for the causal variational principle
in the non-compact setting as developed in~\cite{noncompact}.

From now on, we restrict our attention to non-interacting causal fermion systems in Minkowski space, which we now introduce.

\subsection{Causal Fermion Systems Describing Minkowski Space}\label{sec:continuum}
In this section, we explain how the usual spacetime structures are related to
structures of a causal fermion system in the particular case of the Minkowski vacuum.

We identify Minkowski space with~$\R^{1,3}$, endowed with the standard Minkowski inner product with signature
convention $(+,-,-,-)$, and, for simplicity, choose a fixed reference frame. We consider smooth solutions of the
Dirac equation in Minkowski space, 
\[ 
    (i \gamma^k \partial_k - m) \,\psi = 0 \:, \]
with spatially compact support (i.e.\ with compact support in the spatial variables). The space of solutions is
endowed with the usual scalar product 
\[ ( \psi | \phi ) := \int_{t=\text{const}} (\overline{\psi} \gamma^0 \phi)(t, \vec{x})\, d\vec{x} \:, \]
where $\overline{\psi} = \psi^\dagger\gamma^0$ is the adjoint spinor. As the Hilbert space,~$\H$, of a causal fermion system we choose the completion of the subspace of all negative-energy solutions.
The restriction of the scalar product to $\H$ is denoted by~$\la .|. \ra := (.|.)|_{\H \times \H}$.

The idea is to associate an operator to every point~$x$ of Minkowski space.
For pedagogical reasons, we first explain the construction in the simplified
situation that~$\H$ is replaced by a finite-dimensional Hilbert space generated by a
finite number of continuous wave functions. In this case, the pointwise inner product of the wave
functions defines a bounded sesquilinear form~$b_x$ on~$\H$,
\beq \label{bxdef}
b_x \::\: \H \times \H \rightarrow \C\:,\quad
b_x(\psi, \phi) = - \overline{\psi(x)} \phi(x) \:.
\eeq
(note that, on a finite-dimensional Hilbert space, every sesquilinear form is bounded).
We now represent this sesquilinear form (in the scalar product of the Hilbert space) by a bounded operator, denoted by~$F(x)$, i.e.,
\beq \label{bxrep}
b_x(u, v) = \la u \,|\, F(x)\, v \ra_\H \qquad \text{for all~$u,v \in \H$}\:,
\eeq
where $F(x)$ is uniquely determined by $b_x$.
We combine~\eqref{bxdef} and~\eqref{bxrep} in the form
\begin{equation}\label{eq:correlation}
    \la \psi | {F(x)} \phi \ra = -\overline{\psi(x)} \phi(x) \qquad \text{for all~$\psi, \phi \in \H$} \:.
\end{equation}
The operator $F(x)$ encodes information on the local densities and correlations of the wave functions in~$\H$,
at the spacetime point~$x$. It is referred to as the {\em{local correlation operator}}.
By construction, this operator is selfadjoint, has rank at most four and has
at most two positive and two negative eigenvalues.
Therefore, following Definition~\ref{def:cfs}, the local correlation operator~$F(x)$ is an element of~$\F$,
assuming the spin dimension of $\F$ is given by~$n=2$.
These operators give rise to a mapping,~$F$,
from Minkowski space to $\F$. Defining the measure $\rho$ as the push-forward of the
Lebesgue measure,~$\mu$, of Minkowski space, i.e.,\
\begin{equation} \label{pushforward}
\rho = F_* \mu\,, \qquad \text{or, equivalently,} \qquad
    \rho(\Omega) := \mu \big( {F^{-1}(\Omega)} \big) \:,
\end{equation}
we obtain a causal fermion system~$(\F, \H, \rho)$ of spin dimension two.

The above construction does not apply to our choice of~$\H$ as the space of all negative-energy
solutions of the Dirac equation, because~$\H$ is infinite-dimensional,
and the wave functions in~$\H$ are defined only up to sets of measure zero.
As a consequence, the sesquilinear form in~\eqref{bxdef} is ill-defined.
In order to cure this problem, one needs to introduce an ultraviolet regularization by setting
\begin{equation} \label{psireg}
\psi_\varepsilon={\mathfrak{R}}_\varepsilon(\psi) \:,
\end{equation}
where the {\em{regularization operators}} ${\mathfrak{R}}_\varepsilon : \H \rightarrow C^0(\R^{1,3}, \C^4)$
are linear operators whose range consists of continuous wave functions, and which converge to the identity, as $\varepsilon \searrow 0$, i.e.,
\[ \psi=\lim_{\varepsilon \searrow 0}{\mathfrak{R}}_\varepsilon(\psi) \:. \]
A simple example of a regularization operator is given by convolution with a suitable mollifier.
Working with the regularized wave functions, the right side of~\eqref{eq:correlation}
is a bounded sesquilinear form. Therefore, we can introduce the
{\em{regularized local correlation operator}}~$F^\varepsilon(x)$ by
\[ 
    \la \psi \,|\, {F^\varepsilon(x)} \phi \ra := -\overline{\psi_\varepsilon(x)} \phi_\varepsilon(x)  \qquad \forall \psi, \phi \in \H \:, \]
and applying the above construction to~$F^\varepsilon$ gives a causal fermion system~$(\H, \F, \rho)$.

This construction requires some explanations. First, we note for clarity that the support of the
resulting measure~$\rho$ is given by the image of~$F^\varepsilon$,
\[ M := \supp \rho = \overline{F^\varepsilon(\R^{1,3})} \:. \]
Thus, although~$\F$ is infinite-dimensional (more precisely, the operators in~$\F$ which
have exactly two positive and exactly two negative eigenvalues form an infinite-dimensional Banach manifold),
the support of~$\rho$ is four-dimensional.
It is a general concept that the causal action principle should give rise to measures which ``concentrate''
on low-dimensional subsets of~$\F$. This effect has been studied and proven in simple examples
in~\cite{support, sphere}. This analysis also reveals the underlying mechanisms.

Next, we wish to point out that the causal fermion system introduced above only contains partial information on
the system of Dirac fermions in Minkowski space.
Indeed, restricting attention to~$(\H, \F^{\varepsilon}, \rho)$, some of the structure of Minkowski space, such as its geometry or its causal structure, gets lost. Yet, it is still encoded in the local correlation operators.
Indeed, as is shown in~\cite[Section~1.2]{cfs}, the entire structure of Minkowski space can be recovered
from the causal fermion system, asymptotically, as~$\varepsilon \searrow 0$.
For example, as $\varepsilon$ approaches $0$, the causal structure introduced in Definitions~\ref{def:causalstructure} and~\ref{def:order} reproduces the causal structure of Minkowski space.

The choice of $\H$ as the space of all negative-energy solutions of the Dirac equation realizes Dirac's original proposal that, in the vacuum state of the theory, all the states of negative energy must be occupied ({\em{Dirac sea}}). In the theory of causal fermion systems, this concept is taken seriously. However, the original problems inherent in this concept (like the infinite negative energy density of the sea) do not arise, because the measure describing the Dirac sea introduced above
is a critical point of the causal action principle in the above-mentioned
continuum limit. In simple terms, this means that
the ``states of the Dirac sea drop out of the Euler-Lagrange equations''. As a consequence, measures representing interacting
systems are realized by finite perturbations relative to the sea, giving rise to the usual description
in terms of particles and anti-particles.
The regularization operator $\mathfrak{R}_{\varepsilon}$ has the effect of ``smoothing'' the wave functions on a microscopic scale.
The length scale $\varepsilon$ involved in its definition can be thought of as the Planck length. In CFS theory, the regularization is not merely a technical tool in order to make ill-defined expressions meaningful, but it realizes the idea that, on microscopic length scales, the structure of spacetime must be modified. With this in mind, we  consider the {\em{regularized objects}} as the \textit{physical objects}.

In order to describe systems involving particles and/or anti-particles, following Dirac's hole theory, one extends $\H$ by solutions of the Dirac equation of positive energy and/or removes states of negative energy. Bosonic fields, e.g., the electromagnetic or gravitational fields, correspond to collective ``excitations'' of the Dirac sea and spinorial wave functions described by a Dirac equation modified by a potential~${{\mathcal{B}}}$,
\[ (i \gamma^j \partial_j + {{\mathcal{B}}} - m ) \psi = 0 \:. \]
In order to make this picture precise, one makes use of the fact that
in a subtle analysis of the asymptotics~$\varepsilon \searrow 0$,
referred to as the {\em{continuum limit}}
(for details see~\cite[Chapter~4]{pfp} and~\cite[Section~2.4]{cfs}),
the measure $\rho$ describing the Minkowski vacuum turns out to be a critical point of the causal action.
If particles and/or anti-particles are present, this is no longer the case.
The measures, $\rho$, that are critical points of the causal action functional defined in~\eqref{Sdef}
are then perturbations of the measure describing the Minkowski vacuum state. They lead to a description of interactions among the Dirac particles, which,
asymptotically as~$\varepsilon \searrow 0$, can be described by bosonic fields. As already mentioned above, the analysis sketched here enables one to derive classical field equations, in particular the Maxwell equations and the Einstein field equations, from the causal action principle. Details are presented in~\cite[Chapters~3-5]{cfs}.

\section{Summary of the ``ETH Approach'' to Quantum Theory}\label{sec:ETH}
In this section, we give an overview of the most important concepts underlying the ETH approach,
assuming that most readers are not familiar with this approach to quantum theory, yet. For a recent review, see \cite{froehlich2019review} (see also the survey article \cite{blanchard2016garden}). For technical details on the non-relativistic formulation, see \cite{frohlich2015math}, and, for the relativistic formulation, see \cite{froehlich2019relativistic}.
\subsection{Non-Relativistic ETH} 
We begin our review with a discussion of the non-relativistic formulation of the ETH approach, following closely the presentation in \cite{froehlich2019review}. This approach relies on the \textit{Heisenberg picture} of Quantum Mechanics.
We consider an \textit{isolated open physical system}, $S$. An {\em{isolated}} system is one that has negligible interactions 
with its complement, i.e., with the rest of the universe; an {\em{open}} system $S$ is one that may emit
modes, e.g., electromagnetic radiation, that cannot be monitored by any devices belonging to $S$, because 
they escape from $S$ at the limiting speed ($=\infty$, in the non-relativistic limit, and $=$ speed of light, for relativistic systems). By $\H$ we denote the Hilbert space of pure state vectors of $S$. 
For simplicity, we assume that \textit{all physically relevant states} of $S$ can be described by 
density operators acting on $\H$. We further assume that all physical quantities characteristic 
of $S$ are described by certain bounded selfadjoint linear operators $X=X^*$ acting on $\H$. 
In the Heisenberg picture, these operators evolve in time. For autonomous systems, time evolution of operators in the Heisenberg picture is described by conjugation with a one-parameter group of unitary operators $\big\{U(t)\big\}_{t\in \mathbb{R}}$
defined on the Hilbert space $\H$, referred to as the propagator of $S$.

Operators representing physical quantities of $S$ generate a
$^*$-algebra of operators contained in $\Lin(\H)$, which is abelian for classical systems, but \textit{non-commutative} in QT.
The $^*$-algebra,~$\la A \ra$, generated by a set of operators $A \subset \Lin(\H)$ is the smallest \mbox{$^*$-subalgebra} 
of $\boundedlin (\H)$ satisfying $A\subset \la A \ra$.
Depending on the question under discussion, it is convenient to take completions of $\langle A \rangle$ with respect to
different topologies, the weak, strong or uniform topology. 

Since we are considering non-relativistic systems, time is taken to be absolute, the time axis being given by 
$\mathbb{R}$. For simplicity (or to achieve mathematical precision), one may, however, often assume time to be discrete, with values in $\mathbb{Z}$.

Following \cite{frohlich2015math}, a (non-relativistic) \textit{isolated open system} $S$ is described by a co-filtration, 
$\{\mathcal{E}_{\geq t}\}_{t \in \mathbb{R}}$, of von Neumann algebras, $\mathcal{E}_{\geq t}$, with the property that 
\begin{equation}\label{inclusion}
\mathcal{E}_{\geq t'} \subseteq \mathcal{E}_{\geq t} \quad \text{if   }\, t'>t.
\end{equation}
The algebra $\mathcal{E}_{\geq t}$ is interpreted as the algebra generated by all operators representing \textit{potential events possibly featured by $S$ at or after time} $t$. It is assumed that $\mathcal{E}_{\geq t}$ is closed in the weak topology of the algebra $\boundedlin(\H)$
of all bounded operators acting on $\H$, for all times $t\in \mathbb{R}$. It is expected that, for typical systems, there exists a \textit{universal} (von Neumann) algebra, $\mathcal{N}$, such that
\begin{equation}\label{iso}
\mathcal{E}_{\geq t} \simeq \mathcal{N}, \quad \forall \,t \in \mathbb{R}.
\end{equation}
The algebra $\mathcal{E}$ of all {\em{potential events}} that may happen in the course of history is defined by
\[ 
\boundedlin(\H) \supseteq \mathcal{E}:= \overline{\bigvee_{t \in \mathbb{R}} \mathcal{E}_{\geq t}} \:, \]
where the bar denotes closure in the operator norm of $\boundedlin(\H)$.

In~\cite{froehlich2019review}, potential events possibly featured by $S$ at or after time $t$ are represented by families 
of \textit{disjoint orthogonal projections} acting on $\H$ contained in the algebra $\mathcal{E}_{\geq t}$. 
\begin{Def} A \textit{potential event} possibly featured by $S$ at or after time $t$ is a family 
\[ 
	\big\{\pi_{\xi}\: \vert\: \xi \in \mathfrak{X} \big\}\subset \mathcal{E}_{\geq t}\,, \qquad \mathfrak{X} \text{ a Hausdorff topological space  },
\]
	of disjoint orthogonal projections, $\pi_{\xi}$, on $\H$, with the properties
\[ 
\pi_{\xi}\cdot \pi_{\eta} = \delta_{\xi \eta} \:\pi_{\xi}\ \,,\,\mbox{for all } \xi, \eta \text{ in }\mathfrak{X}\,, \quad\mbox{and}\quad \sum_{\xi \in \mathfrak{X}} \pi_{\xi} = {\bf{1}}\,,
\]
	(i.e., 
	 they form a ``partition of unity'').
\end{Def}

We note that it is assumed here that the projections describing potential events possibly featured by $S$ can be localized in time, the localization being given by a half axis $\mathfrak{I}_{\geq t} := \big\{t'\: \vert\: t' \in \mathbb{R}, t'\geq t \big\}$, and that these projections form the \textit{lattice of projections} associated with the von Neumann algebra $\mathcal{E}_{\geq t}$. For autonomous systems,  potential events,~$\big\{\pi_{\xi}\: \vert\: \xi \in \mathfrak{X} \big\}$, can be translated in time by an amount $t$ by using the propagator $U$ introduced above: Defining 
$$\pi_{\xi}(t):= U(t)\pi_{\xi} U(-t), \quad \text {with } \, U(-t)=U(t)^{*}\,,$$
the family $\big\{\pi_{\xi}(t) \,\vert\, \xi \in \mathfrak{X} \big\}$ is a potential event that arises from the potential event 
$\big\{\pi_{\xi}\: \vert\: \xi \in \mathfrak{X} \big\}$ by translation in time by an amount $t$. If 
$\big\{\pi_{\xi}\: \vert\: \xi \in \mathfrak{X} \big\}$ is localized in $\mathfrak{I}_{\geq t'}$, meaning that this potential event might happen at or after time $t'$, then $\big\{\pi_{\xi}(t) \,\vert\, \xi \in \mathfrak{X} \big\}$ is a potential event localized in 
$\mathfrak{I}_{\geq (t'+t)}$. For the sake of simplicity, we will always assume that the sets $\mathfrak{X}$ labelling the projections representing a potential event are \textit{discrete}.

For autonomous systems with finitely many degrees of freedom, the algebras $\mathcal{E}_{\geq t}$ coincide with 
$\boundedlin(\H)$ and, thus, are \textit{independent} of $t$. It turns out that, for such systems, it is \textit{impossible} to introduce a notion of \textit{events} actually happening at some time $t$ or later, \textit{and the so-called {\bf{measurement problem}} cannot be solved}. But the situation is radically different if one considers systems for which the inclusions in \eqref{inclusion} are {\em{strict}}, which can happen for systems with infinitely many degrees of freedom describing ``massless modes'' that escape to infinity at the limiting speed.

Thus, in order to be able to introduce a notion of \textit{events actually happening at some time~$t$ or later} and to describe how such events can be \textit{recorded} in projective measurements of physical quantities,
the {\bf{``Principle of Diminishing Potentialities''}} (PDP) has been introduced in~\cite{froehlich2019review}\footnote{In earlier work \cite{frohlich2016quest,blanchard2016garden,frohlich2015math} this principle was called ``loss of access to information (LAI).''}:
\begin{equation}\label{PDP}
\boxed{\quad \mathcal{E}_{\geq t'} \subsetneqq \mathcal{E}_{\geq t} \subsetneqq \mathcal{E}, \quad \text{whenever}\,\,\,\, t'>t. \quad}
\end{equation}
In concrete models illustrating the non-relativistic ETH approach with continuous time, PDP is seen to imply that the spectrum of the Hamiltonian generating the propagator of the system is unbounded \textit{above} and \textit{below}; see \cite{F}. (It should be noted, however, that relativistic theories are better behaved in this respect; in particular, the usual spectrum condition can be maintained. This has been shown in \cite{buchholz2014new}.)

A more precise form of \eqref{PDP}, assuming that time is \textit{continuous} is that the relative commutant,
$$\big(\mathcal{E}_{\geq t'}\big){'} \cap \mathcal{E}_{\geq t},\quad \text{with   }\,\, t'>t ,$$
is an infinite-dimensional non-commutative algebra. (We follow the convention that the algebra of all operators on $\H$ commuting with all operators in a subalgebra $\mathcal{A}$ of $\Lin(\H)$ is denoted by $\mathcal{A}'$. If time is \textit{discrete}, the relative commutant considered above can be a finite-dimensional algebra; see \cite{F} for a concrete model.)

To summarize our discussion, an \textit{isolated open system}, $S$, is described, quantum-mechanically, in terms of a co-filtration, $\lbrace \mathcal{E}_{\geq t} \rbrace_{t \in \mathbb{R}}$ (or $\lbrace \mathcal{E}_{\geq t} \rbrace_{t \in \mathbb{Z}}$), of von Neumann algebras, $\mathcal{E}_{\geq t}$, satisfying PDP, all represented on a common Hilbert space $\H$, whose lattices of projections describe potential events. 

Let $\Omega$ be the density matrix on $\H$ representing the actual state of a system $S$. We use the standard notation
$$\omega(X):= \tr(\Omega\,X), \qquad \forall X \in \Lin(\H),$$
to denote the expectation value of an operator $X$ in the state $\omega$ determined by $\Omega$. We define
\begin{equation}\label{state-rest}
\omega_{\,t}(X):= \omega(X), \qquad \forall X \in \mathcal{E}_{\geq t},
\end{equation}
to be the restriction of the state $\omega$ to the algebra $\mathcal{E}_{\geq t}$. As a consequence of~\eqref{PDP}, the restriction, $\omega_{\,t}$, of a state $\omega$ on the algebra
$\mathcal{E}$ to the subalgebra $\mathcal{E}_{\geq t} \subsetneqq \mathcal{E}$ will usually be a \textit{mixed} state \textit{even} if $\omega$ is a pure state on $\mathcal{E}$ (entanglement).

Next, we formulate a criterion that enables one to decide whether an event actually happens at or after time $t$ (or not).\footnote{This criterion is inspired by the desire to rescue as many of the more welcome features of the \textit{Copenhagen interpretation} of quantum mechanics as possible.}
\begin{Def} {\bf{(Centralizer)}} \label{defcentralizer}
Given a $^{*}$-algebra $\mathcal{A}$ and a state $\omega$ on $\mathcal{A}$, the centralizer, $\mathcal{C}_{\omega}(\mathcal{A})$, of the state $\omega$ is the subalgebra of all operators $Y \in \mathcal{A}$ with the property that 
$$\omega([Y, X]) =0, \qquad \forall X \in \mathcal{A},$$
i.e.,
$$\mathcal{C}_{\omega}(\mathcal{A}):=\left\{Y\in \mathcal{A}|\,\,\omega([Y, X]) =0, \,\,\forall X \in \mathcal{A}\right\} .$$
\end{Def}
We note, in passing, that the state $\omega$ defines a finite (normalized) trace on its centralizer 
$\mathcal{C}_{\omega}(\mathcal{A})$. This enables one to classify those von Neumann algebras that could be centralizers of normal states on von Neumann algebras.
\begin{Def} {\bf{(Center of the Centralizer)}}
The \textit{center} of the centralizer $\mathcal{C}_{\omega}(\mathcal{A})$, denoted by 
$\mathcal{Z}_{\omega}(\mathcal{A})$, is the abelian subalgebra of $\mathcal{C}_{\omega}(\mathcal{A})$ consisting
of all operators that commute with \textit{all} other operators in 
$\mathcal{C}_{\omega}(\mathcal{A})$, i.e.,
$$\mathcal{Z}_{\omega}(\mathcal{A}):= \big\{ Y \in \mathcal{C}_{\omega}(\mathcal{A})| \,\,[Y,X]=0\,,\, \forall X \in \mathcal{C}_{\omega}(\mathcal{A}) \big\}. $$
\end{Def}
 We note that the center, $\mathcal{Z}(\mathcal{A})$, of an algebra $\mathcal{A}$ is contained in 
 $\mathcal{Z}_{\omega}(\mathcal{A})$, for all states $\omega$ on $\mathcal{A}$. With these definitions at hand, we define events actually happening in a system $S$ in the following way.
 \begin{Def} {\bf{(Event)}} \label{ETHevent}
Let $S$ be an isolated open system described by a co-filtration $\lbrace \mathcal{E}_{\geq t} \rbrace_{t \in \mathbb{R}}$ of von Neumann algebras. A potential event $\lbrace \pi_{\xi}, \xi \in \mathfrak{X} \rbrace \subset \mathcal{E}_{\geq t}$
(featured by $S$) is happening at time $t$ or later if $\mathcal{Z}_{\omega_t}(\mathcal{E}_{\geq t})$ is non-trivial,
\footnote{The algebra $\mathcal{Z}_{\omega_{t}}(\mathcal{E}_{\geq t})$ is an \textit{abelian} von Neumann algebra.
On a separable Hilbert space, it is generated by a single self-adjoint operator $X$, whose spectral decomposition yields the projections, $\lbrace \pi_{\xi}, \xi \in \mathfrak{X} \rbrace$, describing a potential event. This is the motivation behind \eqref{act-event}.} 
\begin{equation}\label{act-event}
\lbrace \pi_{\xi}, \xi \in \mathfrak{X} \rbrace \,\,\text{generates   }\,\, \mathcal{Z}_{\omega_{\,t}}\big(\mathcal{E}_{\geq t}\big),
\end{equation}
and the ``Born probabilities''
\[ 
\omega_{\,t}(\pi_{\xi_{j}}) \,\,\text{ are \textit{strictly positive}}\,, \,\, \text{ for  }\,\xi_j \in \mathfrak{X}, \,\, j=1,2, \dots, n\,, \]
for some~$n\geq 2$.
 \end{Def}
Next, we have to find out the consequences of the statement that, in an isolated open system $S$, an event actually happens at time $t$ or later. Let 
$\omega_{\,t}$ be the state of~$S$ right before time $t$. Let us assume that an event $\lbrace \pi_{\xi} \,|\, \xi \in \mathfrak{X} \rbrace$ generating $\mathcal{Z}_{\omega_t}(\mathcal{E}_{\geq t})$ happens at time $t$ or later. In the ETH approach, one requires the following axiom (see \cite{froehlich2019review}).
\begin{axiom}\label{collapsaxiom} The actual state  of the system $S$ right after time $t$ is given by one of the states
\[ 
\omega_{\,t, \xi_{*}}(\cdot):=[\omega_{\,t}(\pi_{\xi_{*}})]^{-1}\,\omega_{\,t}\big(\pi_{\xi_{*}} (\cdot) \pi_{\xi_{*}}\big) \]
for some $\xi_{*} \in \mathfrak{X}$ with $\omega_{\,t}(\pi_{\xi_{*}})>0$. The probability for the system $S$ to be found in the state $\omega_{\,t,\xi_{*}}$ right after time $t$ when the event 
$\lbrace \pi_{\xi}, \xi \in \mathfrak{X} \rbrace$ starts to happen is given by Born's Rule, i.e., by
\begin{equation}\label{Born}
\text{\rm{prob}}\{\xi_{*}, t\} = \omega_{\,t}(\pi_{\xi_{*}}).
\end{equation}
\end{axiom}
The ETH approach to QT yields the following picture of the \textit{dynamics of states} in quantum theory: The evolution of states of an isolated open system $S$ featuring events, in the sense of Definition \ref{ETHevent} proposed above, is given by a \textit{stochastic branching process}, whose state space is referred to 
as the {\em{non-commutative spectrum}}, $\mathfrak{Z}_{S}$, of $S$ (see in~\cite{froehlich2019review}). Assuming that~\eqref{iso} holds, the non-commutative spectrum, $\mathfrak{Z}_{S}$ (or $\mathfrak{X}_{S}$), of $S$ is defined by 
\begin{equation}\label{NCspect}
\mathfrak{Z}_{S}:= \bigcup_{\omega} \mathcal{Z}_{\omega}(\mathcal{N})\,, \qquad \text{with  }\,\,\mathfrak{X}_{S}:= \bigcup_{\omega}  \text{spec}\big(\mathcal{Z}_{\omega}(\mathcal{N})\big)\,,
\end{equation}
where the union over $\omega$ is a disjoint union, and $\omega$ ranges over \textit{all} physical states of $S$.
Born's Rule \eqref{Born}, together with ~\eqref{act-event}, specifies the branching probabilities of the process.

To conclude this section, we mention that the states of an isolated open system $S$, so far defined in terms of density operators on the Hilbert space $\H$ associated with $S$, can be viewed, equivalently, as~\textit{additive (probability) measures} on the lattices of orthogonal projections of the algebras $\mathcal{E}_{\geq t}\,, t \in \mathbb{R},$ characterizing the system, i.e., as additive measures on the lattices of potential events possibly happening in $S$.

Let us denote by $\mathcal{P}_{\ge t}$ the lattice of orthogonal projections in $\mathcal{E}_{\ge t}$. As the algebra 
$\mathcal{E}_{\ge t}$ is assumed to be generated by all potential events localized in $\mathfrak{J}_{\ge t}$, the lattice 
$\mathcal{P}_{\ge t}$ can be interpreted as describing all potential events possibly happening at time $t$ or later. It is then natural to view a state on $\mathcal{P}_{\ge t}$ as an assignment of probabilities to all potential events possibly happening at time $t$ or later.

\begin{Def}\label{additivemeasure} A map $\mu$  from $\mathcal{P}_{\geq t}$ to the unit interval $[0,1]$ with the properties
\begin{itemize}
\item[\rm{(i)}]{ $\mu(\pi)\in [0,1] \,,\,\mbox{ for all } \pi \in \mathcal{P}_{\geq t}\,, \text{ and  }\, \mu({\bf{1}})= 1\,;$}\\[-0.5em]
\item[\rm{(ii)}]{if $\big\{\pi_{\xi}\: \vert\: \xi \in \mathfrak{X} \big\}$ is an arbitrary potential event in $\mathcal{P}_{\geq t}$, and 
$\mathfrak{X}_{0} \subseteq \mathfrak{X}$ is an arbitrary subset of $\mathfrak{X}$ then
\[ 
\mu\bigg(\sum_{\xi \in \mathfrak{X}_{0}} \pi_{\xi}\bigg) = \sum_{\xi \in \mathfrak{X}_{0}} \mu(\pi_{\xi})\,,
\]
}
\end{itemize}
is called an \textbf{additive measure} on 
$\mathcal{P}_{\geq t}$.
\end{Def}

Obviously, every state on $\mathcal{E}_{\ge t}$ given in terms of a density operator on $\H$ realizes such a map (simply by restriction to $\mathcal{P}_{\ge t}$). The fact that every additive measure on $\mathcal{P}_{\ge t}$ can be extended to a normal state on $\mathcal{E}_{\ge t}$ is a pillar of the mathematical foundation of QT: \textit{Gleason's theorem} (more precisely, its generalization to von Neumann algebras). 

\begin{Thm}\label{gleason} Assume that the von Neumann algebra $\mathcal{E}_{\geq t}$ does not contain any direct summand isomorphic to the algebra of $2\times 2$ matrices, $\mathbb{M}_{2}(\mathbb{C})$. Let $\mu$ be an additive measure on its lattice, $\mathcal{P}_{\geq t}$, of projections. Then $\mu$ extends to a normal \textit{state} $\omega_{\mu}$ on $\mathcal{E}_{\geq t}$.
\end{Thm} \noindent
For a (sketch of) the proof of this theorem see~\cite{BMW}; details are worked out in~\cite[Theorems 3.3.1 and 3.3.2]{dvurevcenskij}. It represents an important generalization of the classic theorem due to Gleason~\cite{Gleason}.

Axiom 1 and Theorem 3.6 express the intuitively obvious experience that the past is a collection of facts (consisting of all events that have actually happened, i.e., it is ``factual''), while the future is an ensemble of potentialities equipped with an assignment of probabilities for the outcome of events. When an event happens at some time $t$, Nature chooses one of the possible outcomes (described by an orthogonal projection) with a probability as given by Born's Rule applied to the additive measure describing the state of the system at time $t$. Thus, in the ETH approach to QT, the \textit{evolution of states} of isolated open systems in time is {\bf{probabilistic}} -- given by the branching process described above -- while the \textit{time evolution of physical quantities} is described by the {\bf{deterministic}} Heisenberg picture dynamics of linear operators. In the Copenhagen interpretation of quantum mechanics, the change in the state of the system happening whenever the system features an event (as the result of some measurement) is formulated as the projection- or ``collapse'' postulate. In the ETH approach, this postulate has a \textit{natural foundation} in the Principle of Diminishing Potentialities, supplemented by Axiom 1.

A concrete model of an isolated open system satisfying the Principle of Diminishing Potentialities~\eqref{PDP}
to which the above analysis can be applied is studied in~\cite{F}.

For the purposes of the present paper, this is all we need to know about the ETH approach in the non-relativistic setting. Before we proceed to discussing a relativistic version of the ETH approach, we would like to take the
opportunity to point out an \\[0.5em]
{\em{Open problem:}} It would be interesting to extend Huaxin Lin’s theorem on almost commuting
selfadjoint operators~\cite{lin-huaxin} to the setting of pairs, $(\omega, X)$, of a state~$\omega$ on a von Neumann algebra~$\mathcal{E}$ and an operator~$X \in \mathcal{E}$ with the property that~$\text{ad}_X(\omega) := \omega([X,.])$ has a tiny
norm. The question is: Does there exist an operator~$\widetilde{X}$ close in norm to~$X$ and a state $\widetilde{\omega}$ close in norm to $\omega$ such that~$\text{ad}_{\widetilde{X}}(\widetilde{\omega})=0$? If so this would imply that if, in a state $\omega$, an event described by the spectral projections of an operator $X=X^{*}$ ``almost happens'' then, in a state close to $\omega$, an event close to the one described by $X$ \textit{does} happen. It then follows that an event close to the one described by $X$ happens in the state $\omega$.

\subsection{ETH in the Relativistic Setting}
In our tentative formulation of the relativistic form of the ETH approach to QT, we follow  \cite{froehlich2019relativistic}. We only review those aspects that are important for a comparison with CFS theory.  

Instead of defining isolated open systems in terms of a co-filtration $\lbrace \mathcal{E}_{\geq t} \rbrace_{t \in \mathbb{R}}$  of algebras indexed by absolute time, as in the non-relativistic setting, we assume that, with every point $P$ in spacetime, one can associate a von Neumann algebra, $\mathcal{E}_P$, interpreted as the ``algebra of potential events that might happen in the future of $P$''. Spacetime is denoted by $\mathcal{M}$ and is assumed to be a topological space with very little additional structure. It is assumed that all the algebras $\mathcal{E}_P, P \in \mathcal{M},$ are contained in a $C^*$-algebra $\mathcal{E}$. Notions such as centralizers of states on the algebras $\mathcal{E}_P$ and their centers, as well as the corresponding notion of potential events retain the meaning they have in the non-relativistic setting; (it suffices to replace the subscript~``$\ge t$'' by~``$P$'').

Given a state $\omega_{P}$ on the algebra $\mathcal{E}_{P}$, we say that an {\em{event happens in the future of the spacetime point $P$}} if the center $\mathcal{Z}_{\omega_P}$ of the centralizer 
$\mathcal{C}_{\omega_P}(\mathcal{E}_P)$ satisfies the conditions in Definition \ref{ETHevent}.

It is expected that all events that actually happen can be localized in bounded regions of spacetime $\mathcal{M}$. 
If true, this would imply that the operators 
$\big\{ \pi_{\xi} \,\vert\, \xi \in \mathfrak{X}\big\}$ representing a potential event in the future of the point $P$ can 
be localized in a compact region of spacetime contained in what represents the future of $P$ (in a Lorentzian 
spacetime, the future light cone with apex at $P$).
However, at this point, we emphasize that questions regarding the localization in spacetime of events that actually happen have not found compelling answers, so far. One must therefore replace deep understanding by plausible assumptions. For example, one may want to assume that the size of a spacetime region in which an event can be localized is bounded  below by the Planck scale, as suggested by spacetime uncertainty relations.

Leaving these questions aside, the family of algebras $\big\{\mathcal{E}_{P}\big\}_{P\in \mathcal{M}}$ equips spacetime 
$\mathcal{M}$ with a causal structure:

\begin{Def}\label{def:timelikeETH}
A spacetime point $P'$ is in the future of a spacetime point $P$, written as $P'\succ P$ (or, equivalently, $P$ is in the past of $P'$, written as $P\prec P'$\,) if
\begin{equation}\label{PDP2}
\boxed{
\,\mathcal{E}_{P'} \subsetneqq \mathcal{E}_{P} \qquad \text{and} \qquad (\mathcal{E}_{P'})' \cap \mathcal{E}_{P} \,\,\text{is an } \infty-\text{dim. non-commutative algebra}\,.
}
\end{equation}
\end{Def}
The causal structure of $\mathcal{M}$ is determined by the condition~\eqref{PDP2}, which expresses the 
``Principle of Diminishing Potentialities'' (PDP) in the relativistic setting (see~\cite{froehlich2019relativistic};
note that $\succ$ defines a strict partial order relation on $\mathcal{M}$).
This principle is actually a \textit{theorem} in an axiomatic formulation of quantum electrodynamics on four-dimensional Minkowski space proposed by
 Buchholz and Roberts; see \cite{buchholz2014new}. 

Thus, within the ETH approach to relativistic QT, a model of an isolated open physical system $S$ is defined by specifying the data
\[ 
S=\big\{\mathcal{M}, \mathcal{E}, \H, \big\{\mathcal{E}_{P}\big\}_{P\in \mathcal{M}},\succ \big\}\,, \]
where $\mathcal{M}$ is a model of spacetime, $\mathcal{E}$ is a $C^{*}$-algebra represented on a Hilbert space 
$\H$, $\big\{ \mathcal{E}_{P} \big\}_{P \in \mathcal{M}}$ is a family of von Neumann algebras 
contained in $\Lin(\H)$ indexed by points in $\mathcal{M}$, and $\succ$ is the relation between timelike separated points of $\mathcal{M}$ 
induced by PDP~\eqref{PDP2}; see Definition \ref{def:timelikeETH}. 

\begin{Def}\label{def:spacelikeETH}
If a spacetime point $P'$ is neither in the future of a spacetime point $P$ nor in the past of $P$,
we say that $P$ and $P'$ are \textit{space-like separated}. This relation is denoted by
$P \text{\Large{$\bigtimes$}} P'$. \hspace{6.4cm} 
\end{Def}
The description of the evolution of states in the relativistic formulation of the ETH approach is considerably more involved than in the non-relativistic setting, because there is no ordered progression of events. 
We assume that it makes sense to choose an open set $\Sigma$ of points in $\mathcal{M}$ that are space-like separated. If we identify $\mathcal{M}= \R^{1,3}$ with Minkowski space, we may think of $\Sigma$ as a subset of a space-like hypersurface of co-dimension one in $\mathcal{M}$. Since all the algebras 
$\mathcal{E}_{P}$, with $P \in \mathcal{M}$, are assumed to be contained in the $C^{*}$-algebra 
$\mathcal{E}$, the following definition is meaningful,
\begin{equation}\label{algebra}
\mathcal{E}_{\Sigma}:= \overline{\bigvee_{P\in \Sigma} \mathcal{E}_{P}}\,,
\end{equation}
where the closure is taken in the weak topology of $\Lin(\H)$.
A state  $\omega_{\Sigma}$ on the algebra $\mathcal{E}_{\Sigma}$ is a normalized, positive linear functional on $\mathcal{E}_{\Sigma}$.

Next, we consider two space-like separated points $P$ and $P'$ in $\Sigma$. We assume that a state 
$\omega_{\Sigma}$ on the algebra $\mathcal{E}_{\Sigma}$ has been chosen, so that the states 
$\omega_{P}=\omega_{\Sigma} \vert_{\mathcal{E}_P}$ and $\omega_{P'}=\omega_{\Sigma}\vert_{\mathcal{E}_P'}$
are specified, too. We assume that, given $\omega_{\Sigma}$, events happen in the future of $P$ and of $P'$. Let 
$\mathcal{Z}_{\omega_P}$ denote the center of the centralizer of the state $\omega_P$ on $\mathcal{E}_P$. Then  
$\mathcal{Z}_{\omega_P}$ describes an event $\big\{\pi^{P}_{\xi} \,\vert\, \xi \in \mathfrak{X}^{P} \big\}$ happening in the future of $P$. Similarly, let
$\mathcal{Z}_{\omega_{P'}}$ be the abelian algebra describing an event happening in the future of the point $P'$.
We require the following axiom.

\begin{axiom}\label{axiom2}
(Events in the future of space-like separated points commute): Let $P\, \bigtimes \, P'$. Then all operators in $\mathcal{Z}_{\omega_P}$ commute with all operators in $\mathcal{Z}_{\omega_{P'}}$. In particular,
$$\hspace{2.2cm} \big[\pi^{P}_{\xi}, \pi^{P'}_{\eta}\big] =0, \,\,\,\forall\, \xi \in \mathfrak{X}^{P}  \text{ and all }\, \eta \in \mathfrak{X}^{P'}.$$
\end{axiom}

Following~\cite{froehlich2019relativistic}, this axiom may be viewed to be a manifestation of what people sometimes interpret as the
 fundamental \textit{non-locality} of quantum theory: projection operators representing events in the future of two space-like separated points $P$ and $P'$ in spacetime are constrained to commute with each other! This implies exactly what in quantum field theory is understood to express \textit{locality} or Einstein causality.
 
{\bf{ Axiom 1}} is adapted from the non-relativistic formulation of the ETH approach in a natural way.

Next, we assume that some slice $\mathfrak{F}$ in spacetime $\mathcal{M}$ is foliated by (subsets of) space-like hypersurfaces 
$\Sigma_{\tau}$:\,
$\mathfrak{F}:=\big\{\Sigma_{\tau} \,\vert\, \tau \in [0,1]\big\}$, where $\tau$ is a time coordinate 
in the spacetime region filled by 
$\mathfrak{F}$. Let $P$ be an arbitrary spacetime point in the leaf $\Sigma_{1}$, and let the ``recent past'' of $P$, 
denoted by $V^{-}_{P}(\mathfrak{F})$, consist of all points in $\bigcup_{\tau < 1} \Sigma_{\tau}$ that are in 
the \textit{past} of $P$, in the sense specified in Definition \ref{def:timelikeETH} above. We wish to tackle the
the following task: Assume that we know the state 
$\omega_{\Sigma_{0}}$ on the algebra $\mathcal{E}_{\Sigma_0}$ (see~\eqref{algebra}). 
Assuming that Axioms 1 and 2 hold, we propose to determine the state $\omega_{P}$ on $\mathcal{E}_{P}$, for the given point 
$P\in \Sigma_{1}$. Let 
$\big\{P_\iota \,\vert\, \iota \in \mathfrak{I}(\mathfrak{F})\big\}$ denote the subset of points in $V^{-}_{P}(\mathfrak{F})$ in whose future events happen, and let 
$$\big\{ \pi_{\xi_\iota}^{P_\iota} \,\big\vert\, \iota \in \mathfrak{I}(\mathfrak{F})\big\} \subset \mathcal{E}_{\Sigma_0}$$ 
be the \textit{actual events} happening in the future of the points $P_\iota\,, \iota \in \mathfrak{I}(\mathfrak{F})$.  
Here $\mathfrak{I}( \mathfrak{F})$ is a set of indices labelling the points in $V^{-}_{P}(\mathfrak{F})$ in whose future events happen. It is assumed that $\mathfrak{I}( \mathfrak{F})$ is countable.

Next, we define a so-called \textit{``History Operator''}
\begin{equation}\label{History}
H\big(V^{-}_{P}(\mathfrak{F})\big):= \vec{\Pi}_{\iota \in \mathfrak{I}(\mathfrak{F})}\,\, \pi_{\xi_\iota}^{P_\iota}\,,
\end{equation}
where the ordering in the product $\Vec{\Pi}$ is such that a factor $\pi_{\xi_\kappa}^{P_\kappa}$ corresponding to a point $P_{\kappa}$ stands to the right of a factor $\pi_{\xi_\iota}^{P_\iota}$ corresponding to a point $P_{\iota}$ if and only if $P_{\kappa} \prec P_{\iota}$ (i.e., if $P_{\kappa}$ is in the past of $P_{\iota}$). But if $P_{\iota} \bigtimes P_{\kappa}$, i.e., if $P_{\iota}$ and $P_{\kappa}$ are space-like separated, then the order of the two factors is irrelevant, thanks to Axiom 2.

Given the state $\omega_{\Sigma_0}$, the state on the algebra $\mathcal{E}_{P}, \, P\in \Sigma_1$, relevant for predicting whether an event happens in the future of $P$ (as defined before Definition~\ref{def:timelikeETH}) is given by
\[ 
\omega_{P}(X) \equiv \omega_{P}^{\mathfrak{F}}\big(X \big) = \big[\mathcal{N}_{P}^{\,\mathfrak{F}}\, \big]^{-1} 
\omega_{\Sigma_0}\big(H(V^{-}_{P}(\mathfrak{F}))^{*} \,X\, H(V^{-}_{P}(\mathfrak{F}))\big)\,, \,X\in \mathcal{E}_{P}\,, \]
where the normalization factor $\mathcal{N}_{P}^{\,\mathfrak{F}}$ is given by
\[ 
\mathcal{N}_{P}^{\,\mathfrak{F}}= \omega_{\Sigma_0}\big(H(V^{-}_{P}(\mathfrak{F}))^{*} \cdot H(V^{-}_{P}(\mathfrak{F}))\big)\,. \]

The quantities $\mathcal{N}_{P}^{\,\mathfrak{F}}$ can be used to equip the tree-like structure,
in~\cite{froehlich2019review} referred to as the ``non-commutative spectrum'' of $S$, of all possible histories of events in the future of points belonging to the foliation $\mathfrak{F}$ with a probability measure.

For the purposes of this paper, the review presented in this section provides sufficiently many details regarding the ETH approach to QT. We close this section with a few comments:

One may be tempted to think that the ETH approach to QT is similar to the many-worlds interpretation of quantum mechanics. Yet, apart from the fact that it is based on the Heisenberg picture (rather than on the Schr\"{o}dinger picture) of quantum mechanics, the ETH approach has the all-important advantage over the many-worlds interpretation that it introduces an unambiguous, sharp rule that determines the branching of worlds, namely Definition~\ref{ETHevent} of an event. This renders any assumption of existence of many parallel worlds superfluous!
Furthermore, the ETH approach furnishes a precise distinction between future and past: The past consists of all events that have actually happened, while the future refers to all potential events that \textit{might} happen. It thereby eliminates the need to introduce all those  ``excess'' worlds present in the many-worlds interpretation that we do not experience.

To conclude this particular discussion we should stress an \textit{essential fact}: 
In the ETH approach, the time evolution of states of an isolated quantum system is  
\textit{not unitary} (actually, not even linear). This conforms to the experience of people working in a laboratory and carrying 
out measurements. In fact, every successful Stern-Gerlach experiment designed to measure a component of the 
spin of a silver atom confirms that, in the presence of events, in particular in the course of successful measurements of 
physical quantities, 
time evolution of  \textit{states} is \textit{not} given by (deterministic, unitary) Schr\"odinger evolution. It is \textit{non-linear} and 
\textit{stochastic}! Other examples confirming this claim concern the loss of wave-like properties of particles propagating 
through a medium or illuminated by light; e.g., the disappearance of interference patterns in the diffraction of particles at a 
double slit or crystal when these particles are tracked by light scattering. Although the founding fathers of quantum mechanics 
were already keenly aware of the fact that ``measurements'' of physical quantities of a system ``interrupt'' the Schr\"odinger 
evolution of its states, many people appear to ignore and then rediscover this fact every few years. The ETH approach 
to the quantum theory of isolated systems featuring events offers a logically coherent quantum-mechanical description of the 
time evolution of states of isolated systems in terms of a new kind of \textit{stochastic branching process} (whose detailed 
properties should obviously be investigated more closely). The detailed predictions of the ETH approach, which appear to 
be free from internal contradictions or inconsistencies, can, in principle, be tested in experiments and hence are falsifiable, as 
they ought to be.

The fact that, in the ETH approach, the dynamics of states is not unitary appears to eliminate all puzzles and problems related 
to information loss in black holes. It deserves to be emphasized, however, that, in the ETH approach, unitary evolution of 
states is recovered as an \textit{approximation} to the evolution of states of isolated open systems that are not strongly 
entangled with the modes they release to the outside world. In a particular non-relativistic model of a static atom with finitely 
many energy levels coupled to a caricature of the radiation field (studied in a forthcoming paper~\cite{F}),
it can be verified that the evolution of states, as described in the ETH approach, interpolates continuously between unitary 
Schrödinger-Liouville evolution and a Markovian evolution. The precise nature of the evolution of states depends on the 
strength of the entanglement between the atom and the modes of the radiation field that escape to infinity.

\section{Comparing the Structures of the ETH Approach and CFS Theory} \label{sec:comparison}
In Sections \ref{sec:CFS} and \ref{sec:ETH},
we have presented summaries of CFS theory and of the ETH approach to QT, respectively. These two theories have been developed independently and with completely different goals in mind: The aim of CFS theory is to provide a unification of General Relativity with the Standard Model, whereas the ETH approach is designed to provide a consistent completion of quantum theory.
Despite this stark difference in objective, it turns out that the mathematical structures which the two theories are based upon have a lot in common. The purpose of the present section is to give a comparison between the mathematical structures of the two theories and to elucidate where they agree and where there are differences.

\subsection{Foundations of the Two Frameworks}
We begin with a discussion of the basic building blocks in both theories.
\subsubsection{ETH Approach}
A central problem in the ETH approach to QT is to determine all potential states that a system $S$ can occupy in the future of some time $t$ (or in the future of a spacetime point $P$, in the relativistic setting), given its state in the immediate past. The different states the system $S$ can occupy after an event has happened can be labelled by all those projections in the family 
$\lbrace \pi_{\xi} \,|\, \xi \in \mathfrak{X} \rbrace$ describing the event that have a strictly positive Born probability, as explained in Definition \ref{ETHevent} and Axiom 1.

\subsubsection{CFS theory}
The constructions of CFS theory start with the set of potential spacetime point operators $\F$, which consists of all
selfadjoint operators of finite rank with at most $n$ positive and at most $n$ negative eigenvalues, where $n$ is the spin dimension. Note that, since
the operators $x\in \F$ are selfadjoint, every spacetime point operator $x$ determines a set of spectral projections $\pi_{\xi_i}$ on $\H$, where $\xi_i$ is the eigenvector corresponding to the non-zero eigenvalue $\lambda_i$ of $x$. In order to end up with a partition of unity, as required in the definition of a potential event, we need to include the projection onto the kernel of $x$
\[ \pi_\perp :={\mathds{1}} - \sum_{i=1}^{k} \pi_{\xi_i} \]
(where $k\leq 2n$ is the total number of non-zero eigenvalues of $x$). Note that, according to the spectral theorem, we can write 
\[ x=\sum_{i=1}^{k} \lambda_i \,\pi_{\xi_i} + 0 \,\pi_\perp \:. \]
\subsubsection{Discussion}
While these structures emerge from different starting points, they are surprisingly similar. The formalism of the ETH approach is more general than the one of CFS theory. For example, the cardinality of the families $\lbrace \pi_{\xi} \,|\, \xi \in \mathfrak{X} \rbrace$ of projections is always less than or equal to $2n+1$ in CFS theory, while, in the ETH approach, there is no such restriction. 

\subsection{Spacetime and Causal Structure.}
Next, we discuss the way spacetime and its causal structure appear in both theories. 
\subsubsection{ETH}
In the ETH approach, one assumes that spacetime $\mathcal{M}$ is a topological space, with a von Neumann algebra, 
$\mathcal{E}_P$, associated with every point $P\in \mathcal{M}$. The causal structure is then as given in Definitions \ref{def:timelikeETH} and \ref{def:spacelikeETH}.
 
Axiom~\ref{axiom2} imposes further restrictions on the causal structure 
defined on the ``space of events''. It asserts that if two
events happen in the future of two spacelike separated points then the projections describing these 
events must commute. This is a fundamental property expressing ``Einstein causality'' and ensuring 
the consistency of our definition of history operators in \eqref{History}. -- 
The idea underlying Axiom 2 is that if two projections that might represent two putative events happening in the future 
of two space-like separated points, $P$ and $P'$, did \textit{not} commute, then these projections do \textit{not} 
really represent events; their ranges might then be contained in the range of \textit{another projection} 
representing an event that has happened in the future of a spacetime point $P_0$ situated in the \textit{common past} 
of $P$ and $P'$ (i.e., $P_0 \prec P$ \textit{and} $P_0 \prec P'$). 

Clearly, these ideas should be clarified in further studies.

\subsubsection{CFS}
In CFS theory, the structures are introduced in a different order and encode a lot more information.
First, one defines a causal structure on $\F$ by Definition \ref{def:causalstructure}. Then, in order to construct spacetime, one has to find a minimizing measure of the causal action~\eqref{Sdef}. Spacetime is then given by $M:=\supp\rho$, the support of the minimizing measure $\rho$.

The causal structure introduced in Definition \ref{def:causalstructure} allows one to assign a future algebra to every point $x$. There are, however, several different ways in which one can define these future algebras. To avoid confusion we only present the definition best suited for the discussion of the results in Section \ref{sec:results}. (We comment on alternative choices in footnotes.) We first define the timelike future of $x$ and the causal future of $x$.

\begin{Def}\label{cfs:timelikefuture}
For any point $x\in \F$ the {\bf{global timelike future}} is defined by
\[ I^\vee(x):=\{y\in \F  \,|\, \text{ $x$ and $y$ are timelike separated and } \mathcal{C}(x,y)\geq0 \} \]
(see Definition~\ref{def:causalstructure}).
Likewise, the {\bf{global causal future}} is defined by
\[ J^\vee(x):=\{y\in \F  \,|\, \text{ $x$ and $y$ are timelike or lightlike separated and } \mathcal{C}(x,y)\geq0 \} \:. \]
\end{Def} \noindent
These definitions ensure that the point $x$ belongs to $I^\vee(x)$ and to $J^\vee(x)$\footnote{We could modify these definitions by requiring that $\mathcal{C}(x,y)>0$ instead of $\mathcal{C}(x,y)\geq0$. This would be the more natural definition, because it corresponds to the usual view that 
a spacetime point is not in its own future. We point out that the different choices
give rise to the same closure in~\eqref{Evvdef}.}.
Next, we define the future algebras associated with spacetime point operators\footnote{Here again, alternative definitions are possible. First, we need to decide whether we want to take the future algebra to be generated by the timelike future or the causal future. Furthermore, we need to decide whether to take a closure of the algebra, or not.
In case we do take a closure we need to specify the corresponding topology.}.
\begin{Def} {\bf{(Global Future Algebra)}} \label{cfs:causalfuture}
The future algebra of a point $x\in \F$ is given by
\beq \label{Evvdef}
\mathcal{E}^\vee(x)= \overline{\la I^\vee(x) \ra}
\eeq
(where we take the closure in the norm topology of~$\Lin(\H)$).
\end{Def} \noindent

\subsubsection{Discussion}
In general, the causal relations introduced in Definition~\ref{def:order} are not transitive, and hence the statement 
that $y$ is in the (timelike) future of $x$ does not imply that $I^\vee(y)\subset I^\vee(x),J^\vee(y)\subset J^\vee(x) $.
As a consequence, the inclusion $\mathcal{E}^\vee(y)\subset \mathcal{E}^\vee(x)$ does not hold in general.
Clearly, following the ideas of the ETH approach,
one could consider the inclusion $\mathcal{E}^\vee(y)\subset \mathcal{E}^\vee(x)$ to define a new causality 
relation on $\F$. This would have the advantage that the causal relations would become transitive.
However, this procedure looks like an ad-hoc method to enforce an analogy between the structures
of CFS theory and the ETH approach.
At this point, we mention that another notion of causality
can be introduced on a given causal fermion system that is transitive, starting from analytic properties of
linearized field equations (which describe linear perturbations of a minimizing measure; see~\cite[Section~4.1]{linhyp}).
The precise connections between these different definitions of causal relations
have not been explored, yet. It seems that the notion of causality introduced in
Definition~\ref{def:causalstructure} might be more fundamental than other notions, because 
it reflects the structure of the causal Lagrangian~\eqref{Lagrange}.

\subsection{The Role of States}
When studying a physical system, the definition of its space of states plays a crucial role.
\subsubsection{ETH}
In the ETH approach, states $\omega_P$ are defined as additive measures on the lattice of orthogonal projections in the von Neumann algebra $\mathcal{E}_P$, or, equivalently, as normal states on $\mathcal{E}_P$. States play two roles:
\bitem
\item[(i)] When determining whether an actual event takes place in the future of a spacetime point $P$, knowledge 
of the state $\omega_P$ plays a crucial role, in that an event is represented by a family of projections belonging to 
the center, $\mathcal{Z}_{\omega_P}$, of the centralizer, 
$\mathcal{C}_{\omega_P}(\mathcal{E}_P) \subset \mathcal{E}_P$, of the state $\omega_P$. 
\item[(ii)]
When an actual event takes place, the state of the system in the immediate past of the event determines the 
(Born) probabilities of the different possible outcomes of the event and, hence, of the different states the system 
could occupy after the event has happened. According to Axiom 1, Nature will then choose one of the states that has
a positive Born probability.
\eitem

\subsubsection{CFS}
In CFS theory, the state of the physical system is determined globally in spacetime by prescribing the measure~$\rho$.
In this sense, this measure plays a similar role as the state in the conventional Heisenberg picture
of QT. To some degree, this analogy extends to the role of the state in the ETH approach,
where it is used to form the centralizer of an algebra,
a subalgebra whose center determines the events (see Definitions~\ref{defcentralizer}--\ref{ETHevent}).
Similarly, in CFS theory, specifying~$\rho$ gives rise to subalgebras of the future algebras,
namely the algebras generated by the operators that belong to the support of $\rho$ (i.e., are realized in the physical system). More precisely, these algebras are defined as follows.
\begin{Def} {\bf{(Spacetime Restricted Timelike Future)}} \label{cfs:spacetimetimelikefuture}
For any point $x\in M$, the spacetime restricted timelike future is defined by
\[ I_\rho^\vee(x):= I^\vee(x) \cap M = \{y\in M\,|\, \text{ $x$ and $y$ are timelike separated and } \mathcal{C}(x,y)\geq0 \} \]
(see Definition~\ref{def:causalstructure}).
\end{Def}
\begin{Def} {\bf{(Spacetime Restricted Future Algebra)}} \label{defsrfa}
The spacetime restricted future algebra of a point $x\in M$ is given by 
\[ \mathcal{E}_{\rho}^\vee(x) = \overline{\la I_\rho^\vee(x) \ra} \]
(where the closure is taken again in the norm topology of~$\Lin(\H)$).
\end{Def}
This definition is in harmony with the analytical results in Minkowski space presented
in Section~\ref{sec:results}. Indeed,
for causal fermion systems in Minkowski space, the causal relations do become transitive 
in limiting case~$\varepsilon \searrow 0$, i.e., if the regularization is removed.

We remark that the connection to states being density operators acting on Fock spaces
has been unravelled in a recent paper~\cite{fockbosonic}.
Let~${\mathcal{B}}$ be the set of all measures that satisfy
the Euler-Lagrange equations derived from the causal action principle
in the continuum limit (i.e.,\ in the limiting regime~$\varepsilon \searrow 0$).
A measure~$\rho_\text{vac}$ describing the Minkowski vacuum state, 
as constructed in Section~\ref{sec:continuum}, belongs to~${\mathcal{B}}$.
In~\cite{fockbosonic} it is shown at the level of formal power series
(i.e., in a perturbative treatment well-defined to every order, thanks to the
ultraviolet regularization that is built in by the operator $\mathfrak{R}_{\varepsilon}$)
that if~$\rho$ is a ``finite perturbation'' of~$\rho_\text{vac}$
(i.e.,\ if it lies in some neighborhood of~$\rho_\text{vac}$ in~${\mathcal{B}}$),
then it can be described by a state in a bosonic Fock space.
Moreover, in the so-called {\em{holomorphic approximation}}, the dynamics of a causal fermion system
can be described by a one-parameter group of unitary time evolution operators (propagators) acting 
on the Fock space.
Justifying the various approximations involved in this analysis and extending the constructions so as to include 
fermionic Fock spaces is the objective of ongoing research.

\subsubsection{Discussion}
The definition of states in the ETH approach is the standard definition used in QT. However, the characterization of isolated open systems in terms of descending filtrations of algebras of potential events is very unconventional. It leads to a \textit{new stochastic law} determining the \textit{time evolution of states} in the ETH approach! 
Instead, in CFS theory the physical system is described in terms of the space ~$\F$ and its state by a measure
~$\rho$ on the space of spacetime operators.
Despite this major difference,
there is an analogy regarding the role of operator algebras, in that specifying the state of a physical system
(by choosing a state~$\omega$ on a ``future algebra'' $\mathcal{A} \in \mathcal{E}$, the measure~$\rho$, respectively,)
determines a subalgebra of the future algebra
(the centralizer~$\mathcal{C}_{\omega}(\mathcal{A})$, the spacetime restricted
algebra~$\mathcal{E}_{\rho}^\vee(x)$, respectively),
which determines the events that actually happen (via its center, its spacetime point operators, respectively).
In CFS theory, a close connection between the measure~$\rho$ and
conventional states on Fock spaces has been obtained in~\cite{fockbosonic}.
But this is ongoing research, and it is too early to tell whether this will also tighten the
connection to the role of the state in the ETH approach.

\subsection{Dynamical Evolution of States} \label{secdes}
Last but not least we discuss the time evolution of states of systems in both theories. 
\subsubsection{ETH}

If an event happens at or after time $t$, in the sense of Definition~\ref{ETHevent}, then the state of the system 
changes in accordance with the actual outcome of the event, as specified in Axiom~\ref{collapsaxiom}. Quantum theory only predicts the probabilities of different outcomes, but not the actual outcome. Its dynamical law for the evolution of states is therefore {\bf{probabilistic}}, \textit{not} deterministic. The state corresponding to the actual outcome of the event must then be used for further predictions. 
This law of evolution of states of isolated open systems corresponds to a novel type of 
\textit{stochastic branching process} whose state space is the non-commutative spectrum 
introduced in~\eqref{NCspect}.

In the relativistic setting, the dynamical law is more involved, because, besides the fact that the relevant algebras are now labelled by spacetime points rather than just by time, the history operators \eqref{History} have to be determined.

To conclude this discussion, we remark that, in the ETH approach, isolated systems featuring events are necessarily 
systems with infinitely many degrees of freedom. Furthermore, measurements or observations are special types of events 
described by partitions of unity by orthogonal projections that are close to the spectral projections of a selfadjoint operator 
representing the physical quantity that is measured.

\subsubsection{CFS} \label{secdescfs}
When considering variations of the measure~$\rho$ in the causal action principle,
this measure and the spacetime it determines are varied as a whole.
It is therefore not straightforward to formulate a dynamical law in the form of an initial value problem.  

For the linearized field equations (which describe linear perturbations of a minimizing measure),
the Cauchy problem has been formulated in a recent paper~\cite{linhyp}, and well-posedness is proven
for initial data given on a surface layer,
under suitable assumptions, which are subsumed in the so-called {\em{hyperbolicity conditions}}.
Well-posedness implies that, given initial data, the
time evolution is unique.

\subsubsection{Discussion} \label{secdesdiscuss}
The well-posedness of the Cauchy problem for the linearized field equations in~\cite{linhyp}
means that the time evolution is deterministic. 
This is in stark contrast to the probabilistic nature of the evolution of states in the ETH approach.
However, one should keep in mind that the results in~\cite{linhyp}
were obtained only in the linearized approximation
and only if the hyperbolicity conditions hold. For general causal fermion systems
or if one goes beyond the linearized approximation, it is not to be expected that the
Euler-Lagrange equations of the causal action principle allow for a formulation in terms of a
Cauchy problem (see~\cite{cauchy} for an attempt in this direction and a more detailed explanation of the
non-uniqueness).
This leaves open the possibility that the nonlinear dynamics of a general causal fermion system
might be compatible with the stochastic branching process appearing in the ETH approach.
We will come back to this point in Section~\ref{sec:outlook}.


\subsection{Localization of Events in Spacetime}
There is another difference between the ETH approach and CFS theory that has not been drawn attention to, yet.
We have argued that the spacetime point operators $x \in M$ of CFS theory play 
a role similar to the one played by events in
 the ETH approach. By definition, these operators are associated with
individual spacetime points. However, because an ultraviolet regularization is necessarily involved,
 as explained in the example of the Minkowski vacuum in Section~\ref{sec:continuum}; cf.~\eqref{psireg},
one can think of the spacetime operators $x$ as being localized in regions 
of spacetime whose extension is given by the Planck scale.
The operators representing events in the ETH approach are, however, typically localized in spacetime regions 
whose size can be much larger than the Planck scale. If an event is 
localized in a tiny region of spacetime, it is accompanied by large energy fluctuations, as predicted by the 
uncertainty relations. Taking into account gravitational effects, this implies that the extension of events in spacetime
cannot be too small on the Planck scale, because, otherwise, microscopic black holes would form, 
as has been argued in~\cite{doplicher}.
With these observations in mind, we conclude that the spacetime points of a causal fermion system could
only represent a special class of events as defined in the ETH approach.
Once might think of them as those events that are {\em{best (optimally) localized}} in spacetime. Moreover, spacetime point operators in CFS theory can only represent events with small
(namely, at most $2n+1$) number of possible outcomes.

CFS theory might, however, also describe events, in the sense of the ETH approach, whose extension in spacetime exceeds the Planck scale. 
    Indeed, the spacetime restricted future algebra $\mathcal{E}_{\rho}^\vee(x)$
    contains operators of the form 
\[ \phi=\int_M f(x)\,x\, d\rho(x)\,,\]
    with $\supp(f)\subset I_\rho^\vee(x)$, as well as linear combinations of
    products of such operators. These operators are not sharply localized in spacetime. Their spectral decomposition may
    give rise to projections representing potential events, in the sense of the ETH approach, localized in arbitrarily large spacetime regions.

\section{Results for Causal Fermion Systems Describing Minkowski Space}\label{sec:results}
In the previous sections, a connection has been made between spacetime point operators of causal fermion systems and operators representing events in the ETH approach to quantum theory.
In the following, we make this connection more precise by putting it into
the context of results obtained in a recent paper~\cite{neumann},
where operator algebras are studied for causal fermion systems in Minkowski space.
We first state these results and then interpret them.
We consider the causal fermion system constructed in Section~\ref{sec:continuum}, with a regularization operator 
$\mathfrak{R}_\varepsilon$ chosen as the operator of multiplication by a
smooth cutoff function,~$\mathfrak{G}_\varepsilon$, in momentum space (for more details see \cite[Section 2.6]{neumann}). Thus~$\H$ is the Hilbert space of all negative-energy solutions of the free Dirac equation on Minkowski space, and~$\rho$ is the push-forward of the Lebesgue measure of Minkowski space under~$F^\varepsilon$ (see~\eqref{pushforward}). 

Given an open set~$\Omega \subset \R^{1,3}$ of Minkowski space, we 
consider all local correlation operators in~$\Omega$
\[ {\mycal X}_\Omega :=\{F^\varepsilon(x)\:|\: x\in\Omega \}  \subset \Lin(\H) \:. \]
For technical reasons, it is preferable to ``smear'' these local correlation operators by 
smooth functions with compact support in $\Omega$,
\[ 
A^\varepsilon_f:=\int_{\R^{1,3}}f(x)\,  F^\varepsilon(x)\, d^4x\in\Lin(\H),\quad f\in C_0^\infty(\Omega,\C) \:. \]

\begin{Def}\label{algebramink} The (non-unital) $*$-algebra
	\[ \A^\varepsilon_\Omega := \la \{ A^\varepsilon_f \:|\: f\in C_0^\infty(\Omega,\C) \} \ra \]
	is referred to as the {\bf{local algebra}} associated with $\Omega$.
\end{Def} \noindent
The uniform closure of this algebra contains all the operators of~${\mycal X}_\Omega$.
Both sets of operators generate the same $C^*$-algebra,
\begin{equation}\label{formulaclosure} \overline{\A^\varepsilon_\Omega} = \overline{\la {\mycal X}_\Omega \ra} \,,\:
\end{equation}
where the closure is taken in the norm topology of~$\Lin(\H)$. 

In order to analyze connections with the ETH approach, we first show that
the commutator of~$F^\varepsilon(x)$ with elements of~$\mycal{A}_\Omega^\varepsilon$
nearly vanishes if~$\Omega$ lies in the open future light cone centered at~$x$.
The best one may hope to show is that this commutator is tiny in the sense that it
tends to zero as~$\varepsilon \searrow 0$; more precisely,
\beq \label{qualitative}
\big\Vert \,\big[ F^\varepsilon(x), A \big]\, \big\Vert = \O(\varepsilon^p)\, \Vert F^\varepsilon(x) \Vert\: \Vert A \Vert, \qquad \text{for all~$A \in \mycal{A}_\Omega^\varepsilon$} \:,
\eeq
for some exponent $p>0$.
The factor~$\|F^\varepsilon(x)\|$ is needed in order for the estimate to be
invariant under scalings of~$F^\varepsilon$.
In attempting to give this relation a precise meaning, one must keep in mind that the definition of the local algebras 
also depends on~$\varepsilon$. This makes it necessary to relate the algebras associated with different values of
~$\varepsilon$ to one another. For this purpose, we introduce the free algebra~$\A(C^\infty_0(\Omega))$
generated by $C^\infty_0(\Omega)$ and introduce a linear map
\[ 
\iota^\varepsilon : \A(C^\infty_0(\Omega)) \rightarrow\mycal{A}_\Omega^\varepsilon\,, 
\quad \text{defined on monomials by} \quad
f_{i_1}\cdots f_{i_k} \mapsto A_{f_{i_1}}^\varepsilon \cdots A_{f_{i_k}}^\varepsilon \:. \]
Now we may quantify~\eqref{qualitative} as follows (see~\cite[Theorem~4.15]{neumann}).

\begin{Thm} \label{prpiotaes}
	Let $\Omega$ be contained in the open future light cone centered at~$x$. Then,
	for an arbitrary operator $a\in \A(C^\infty_0(\Omega))$, there is a constant $c(a)>0$ such that
	\beq \label{Fiota}
	\big\|\, \big[F^\varepsilon(x),\iota^\varepsilon(a) \big]\,\big\|\le c(a)\, \varepsilon^{\frac{3}{2}}\,\|F^\varepsilon(x)\| \:, \,\, \forall \,\varepsilon>0\,.
	\eeq
\end{Thm} \noindent
We remark that, applying Huaxin Lin's theorem on almost commuting selfadjoint operators~\cite{lin-huaxin},
the estimate~\eqref{Fiota} shows that in small neighborhoods of~$F^\varepsilon(x)$
and~$\iota^\varepsilon(a)$ there are operators which commute.

We also mention another result from~\cite{neumann} that will be important
for the discussion of the PDP~\eqref{PDP2}, in point (ii), below.
For a spacetime point $x \in\R^{1,3}$, the image of the operator~$F^\varepsilon(x)$, denoted by
\[ S_x := \text{range}\, F^\varepsilon(x) \subset \H \:, \]
is a four-dimensional complex vector space
(this is the {\em{spin space}}~\eqref{Sxdef}; for more details see~\cite[Section~1.1.4]{cfs}).
The vectors in the image of~$S_x$ can be written as
\begin{equation}\label{localizedvectors}
P^\varepsilon(\,\cdot\,,x)\chi:\R^{1,3}\ni y\mapsto\int_{\R^4}\frac{d^4}{(2\pi)^4} \:\mathfrak{G}_\varepsilon(k)\:
\delta(k^2-m^2)\:\Theta(-k_0) \:(\slashed{k}+m) \:\chi\, e^{-ik\cdot(y-x)}\,,
\end{equation}
with~$\chi \in \C^4$. These are regular solutions of the Dirac equation, but they develop singularities on the
null cone centered at $x$ in the limit $\varepsilon\searrow 0$.  They can be understood as wave packets
which, at time~$x^0$, are well ``concentrated'' at the spatial point~$\vec{x}$.
We note that it is impossible to localize such wave packets in a compact region,
because such a localization is incompatible with our restriction to negative-energy solutions
(as is made precise in a general setting by Hegerfeldt's theorem~\cite{hegerfeldt1974remark};
a nice proof in the context of the Dirac equation can be found in~\cite[Section~1.8.4]{thaller}).

The following theorem holds (see~\cite[Proposition~4.14]{neumann}).
\begin{Thm}\label{theoremexpvalue}
	Let $\Omega$ be contained in the open future light cone centered at~$x$. Then
	for any $a\in \A(C^\infty_0(\Omega))$ there is a constant $c(a)>0$ such that, for all spinors~$\chi,\tilde{\chi}$ in $\C^4$ and for all~$\varepsilon>0$,
	$$
	|\langle P^\varepsilon(\,\cdot\,,x)\, \chi\,\big|\, \iota^\varepsilon(a) P^\varepsilon(\,\cdot\,,x) \,\tilde{\chi}\rangle |\le c(a)|\chi||\tilde{\chi}|.
	$$
	Assume that the set $\Omega$ intersects the null cone centered at $x$. Let $x_0$ belong to this intersection. Then for every sufficiently small $r>0$ there are~$f\in C^\infty_0(B_r(x_0),\C)$, spinors $\chi,\tilde{\chi}\in\C^4$ and a constant $c> 0$ such that, for sufficiently small $\varepsilon>0$,
	$$
	|\langle P^\varepsilon(\,\cdot\,,x)\, \chi\,\big|\, A_f^\varepsilon P^\varepsilon(\,\cdot\,,x)\, \tilde{\chi}\rangle |\ge \frac{1}{c\,\varepsilon^2} \:.
	$$
\end{Thm} \noindent
This result allows us to detect the null cone centered at the spacetime point~$x$ in Minkowski space by looking at expectation values of elements of the local algebras.

We now discuss these results in relation with the considerations in Section~\ref{sec:comparison}.
In order to relate our results to the spacetime restricted future algebra~$\mathcal{E}_{\rho}^\vee(x)$  defined in Definition~\ref{defsrfa} and taking into account \eqref{formulaclosure}, 
it is most natural to choose the open set~$\Omega$ as the open future light cone centered at $x$ in Minkowski space,
which we here denote by~$\mathcal{I}^\vee(x)$, and to generate its algebra, as in Definition \ref{algebramink},
\[ 
{\mathcal{E}}_M^{\varepsilon,\vee}(x) = \overline{\A^\varepsilon_{\mathcal{I}^\vee(x)}} \]
with the closure taken in the norm topology of~$\Lin(\H)$, where the subscript $M$  is added in order to 
distinguish this algebra from the one constructed in Definition~\ref{defsrfa}, while the superscript $\varepsilon$ 
indicates the dependence on the
regularization. We stress that this algebra is generated by the spacetime point
operators~$F^\varepsilon(x)$ (see \eqref{formulaclosure}) which belong to the support of the measure 
$\rho$ introduced above. Therefore, it is an analog
of the {\em{spacetime restricted future algebra}}~$\mathcal{E}^\vee_{\rho}(x)$
introduced in Definition~\ref{defsrfa} (but not of the global future algebra of Definition \ref{cfs:causalfuture}).

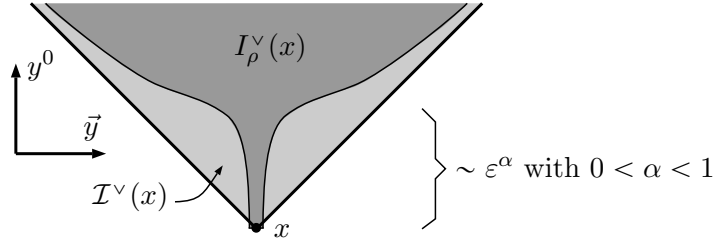
\begin{figure}
	%
	\psscalebox{1.0 1.0} 
	{
		\begin{pspicture}(-4,-1.5997621)(16.351667,1.5997621)
		\definecolor{colour0}{rgb}{0.8,0.8,0.8}
		\definecolor{colour1}{rgb}{0.6,0.6,0.6}
		\pspolygon[linecolor=colour0, linewidth=0.02, fillstyle=solid,fillcolor=colour0](0.24222222,1.5722866)(3.2155557,-1.383269)(6.1977777,1.576731)
		\pspolygon[linecolor=colour1, linewidth=0.02, fillstyle=solid,fillcolor=colour1](0.42,1.5756199)(6.005,1.5756199)(5.935,1.5071625)(5.86,1.433439)(5.77,1.3544496)(5.68,1.2754604)(5.55,1.1804072)(5.405,1.0695561)(5.255,0.97450286)(5.165,0.8902476)(5.02,0.79099226)(4.835,0.67407733)(4.695,0.574822)(4.515,0.48397097)(4.28,0.41918373)(4.08,0.33599225)(3.885,0.2620029)(3.705,0.15721564)(3.585,0.06742841)(3.495,-0.03102904)(3.445,-0.12555031)(3.415,-0.23586947)(3.37,-0.37805033)(3.345,-0.53049713)(3.315,-0.672678)(3.305,-0.85198647)(3.295,-1.0249652)(3.3,-1.3924121)(3.22,-1.3943801)(3.13,-1.3900183)(3.135,-0.97884816)(3.115,-0.7306035)(3.1,-0.55523115)(3.06,-0.39384818)(3.02,-0.25166735)(2.955,-0.08342265)(2.82,0.1237582)(2.685,0.19248159)(2.5,0.29673693)(2.315,0.3604603)(2.105,0.4236518)(1.96,0.4818433)(1.79,0.5608327)(1.455,0.7575348)(1.17,0.973705)(0.88,1.1854072)(0.63,1.3860455)(0.515,1.4808327)
		\pscircle[linecolor=black, linewidth=0.02, fillstyle=solid,fillcolor=black, dimen=outer](3.2155557,-1.4004912){0.065}
		\psline[linecolor=black, linewidth=0.04, arrowsize=0.05291667cm 2.0,arrowlength=1.4,arrowinset=0.0]{->}(0.02,-0.4143801)(0.02,0.7856199)
		\psline[linecolor=black, linewidth=0.04, arrowsize=0.05291667cm 2.0,arrowlength=1.4,arrowinset=0.0]{->}(0.02,-0.4143801)(1.22,-0.4143801)
		\rput[bl](0.92,-0.2393801){\normalsize{$\vec{y}$}}
		\psline[linecolor=black, linewidth=0.04](3.22,-1.4143801)(0.22,1.5856199)
		\psline[linecolor=black, linewidth=0.04](3.22,-1.4143801)(6.22,1.5856199)
		\rput[bl](3.4516666,-1.5238246){\normalsize{$x$}}
		\rput[bl](0.17,0.5656199){\normalsize{$y^0$}}
		\psbezier[linecolor=black, linewidth=0.02](3.305,-1.4043801)(3.3092313,-0.8736793)(3.3308089,-0.118341394)(3.61,0.09561989678276973)(3.8891912,0.3095812)(4.139231,0.34132072)(4.53,0.5006199)(4.9207687,0.6599191)(5.82,1.3856199)(6.02,1.5856199)
		\psbezier[linecolor=black, linewidth=0.02](3.125,-1.4043801)(3.1207688,-0.8736793)(3.104191,-0.118341394)(2.825,0.09561989678276973)(2.5458088,0.3095812)(2.2957687,0.34132072)(1.905,0.5006199)(1.5142313,0.6599191)(0.615,1.3856199)(0.415,1.5856199)
		\rput[bl](2.922778,0.7495088){\normalsize{$I^\vee_\rho(x)$}}
		\rput[bl](1.0433333,-1.183269){\normalsize{${\mathcal{I}}^\vee(x)$}}
		\psline[linecolor=black, linewidth=0.02](3.113889,-1.4043801)(3.316111,-1.4043801)
		\psbezier[linecolor=black, linewidth=0.02, arrowsize=0.05291667cm 2.0,arrowlength=1.4,arrowinset=0.0]{->}(2.1355555,-1.05438)(2.4822223,-1.0454912)(2.5133333,-0.943269)(2.771111,-0.5699356587727845)
		\psline[linecolor=black, linewidth=0.02](5.42,0.18561989)(5.597778,0.007842119)(5.597778,-0.48104677)(5.731111,-0.6143801)(5.597778,-0.74771345)(5.597778,-1.2810467)(5.42,-1.4143801)
		\rput[bl](5.8516665,-0.72382456){\normalsize{$\sim \varepsilon^\alpha$ with $0<\alpha<1$}}
		\end{pspicture}
	}
	\caption{The spacetime restricted causal future~$I^\vee_\rho(x)$ of the causal fermion system
		and the future light cone~$\mathcal{I}^\vee(x)$ of Minkowski space.}
	\label{figL1}
\end{figure}

It is an important difference in the construction of these algebras that ~$\mathcal{E}_{\rho}^\vee(x)$ is defined by using the causal structure of the causal fermion system
(as given by the spacetime restricted timelike future; see Definition~\ref{cfs:spacetimetimelikefuture}),
whereas~${\mathcal{E}}_M^{\varepsilon,\vee}(x)$ involves the causal structure on Minkowski space.
This difference becomes irrelevant asymptotically, as~$\varepsilon \searrow 0$, because, in this
limiting regime, the causal structures agree (for details see~\cite[Section~1.2.5]{cfs}).
However, they do {\em{not}} agree for positive values of the regularization length $\varepsilon$.
At the scale of the regularization length, the spacetime restricted timelike future~$I^\vee_\rho(x)$ of the causal fermion system has a rather complicated form (for details see~\cite{reg} and~\cite[Appendix~A]{jacobson}).
The qualitative behavior is shown in Figure~\ref{figL1}.
Away from the point~$y=x$, the spacetime restricted causal future~$I^\vee_\rho(x)$ is contained strictly inside
the light cone~$\mathcal{I}^\vee(x)$.
The reason is that the causal action principle aims at making the spacelike region as large as possible.
In particular, the singularities of the kernel of the fermionic projector on the Minkowski null cone
are contained in the spacelike region (this is essentially used in the continuum limit analysis in~\cite{cfs}).
The length scale of the deviation of the light cones is~$\sim \varepsilon^\alpha \, m^{\alpha-1}$ with~$0 < \alpha <1$;
thus it is much {\em{larger}} than the regularization length.
All points~$y$ in a small neighborhood of~$x$, however, are timelike separated (in the sense of
Definition~\ref{def:causalstructure}). Therefore, near~$y=x$, the spacetime restricted causal
future~$I^\vee_\rho(x)$ contains
$\mathcal{I}^\vee(x)$. Here the relevant length scale is~$m \varepsilon^2$; thus it is much {\em{smaller}}
than the regularization length.

In~\cite{neumann}, in view of this rather complicated structure of the set~$I^\vee_\rho(x)$, 
we work with the light cones $\mathcal{I}^\vee(x)$ of Minkowski space.
This simplification is admissible, because the deviations of the different light cones
vanish in the limit~$\varepsilon \searrow 0$;
(furthermore, in view of the right side of~\eqref{Fiota}, it is
impossible to avoid error terms, which vanish as~$\varepsilon \searrow 0$).

Theorem~\ref{prpiotaes} shows that the operator~$F^\varepsilon(x)$ commutes with
the algebra in its open future light cone, up to small corrections involving the regularization length.
As a matter of fact, the estimate applies merely to the subalgebra  
$$
{ {\mathcal{W}}^{\varepsilon,\vee}(x)}:=\A^\varepsilon_{\mathcal{I}^\vee(x)}\subset\mathcal{E}_M^{\varepsilon,\vee}(x) \:.
$$
Nevertheless, this subalgebra is dense in ${\mathcal{E}}^{\varepsilon,\vee}_M(x)$ in the norm topology, 
which means that
the difference between these two algebras is more of technical nature, but does not seem to have a
physical significance.

We now discuss a few important consequences of the above estimates related to two of the main concepts underlying the ETH approach: the PDP~\eqref{PDP2} and the notion of an event that actually happens.\\[-0.4em]
\begin{itemize}[leftmargin=2em]
	\item[{\rm{(i)}}] 
	\noindent The estimate of Theorem~\ref{prpiotaes} can be understood as a \textit{loss of access
	to information}. In order to explain the connection, we consider the situation that
	$\mathcal{I}^\vee(\tilde{x})$ is chosen as the future light cone of a point~$\tilde{x}$ in the time-like future of~$x$
	(see Figure~\ref{figL2}).
	Then for any point~$z$ in the open diamond $\mathcal{I}^\vee(x) \cap \mathcal{I}^\wedge(\tilde{x})$,
	the wave functions in the image of the operator $F^\varepsilon(z)$ (given by \eqref{localizedvectors}) are wave packets
	which at time $z^0$ are as far as possible ``concentrated'' at the spatial point~$\vec{z}$. As a consequence, in the limit $\varepsilon \searrow 0$
	these wave functions become singular on the null cone centered at~$z$.
	In other words, the wave functions have a significant high-frequency component which
	propagates almost with the speed of light. The information carried by this component of the wave
	is no longer accessible by the operators in the algebra ${\mathcal{W}}^{\varepsilon,\vee}(\tilde{x})$. 
	The estimate~\eqref{Fiota} can be understood as a quantitative version of this argument.
	Indeed, the factor $\varepsilon^\frac{3}{2}$ comes about by estimating the norm of the
	wave propagating along the arrows in Figure~\ref{figL2}.

We remark for clarity that, in view of spatial localization and frequency splitting as made precise
by Hegerfeldt's theorem, the above wave functions of negative frequency always have ``tails''
which do not propagate almost with the speed of light.
Moreover, as we consider massive fields, the wave will involve components which
propagate with subluminal speed.
However, as is made precise by the above estimates,
by decreasing~$\varepsilon$ the contribution by the tails and by the subluminal contributions
can be made arbitrarily small compared to the high-frequency contribution. \\[-0.3em]
	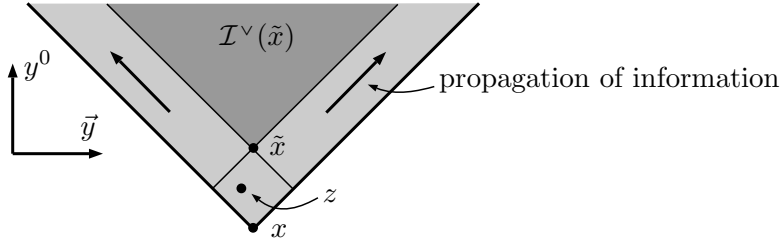
\begin{figure}
		%
		\psscalebox{1.0 1.0} 
		{
			\begin{pspicture}(-3,-1.5997621)(14.467778,1.5997621)
			\definecolor{colour0}{rgb}{0.8,0.8,0.8}
			\definecolor{colour1}{rgb}{0.6,0.6,0.6}
			\pspolygon[linecolor=colour0, linewidth=0.02, fillstyle=solid,fillcolor=colour0](0.24222222,1.5722866)(3.2155557,-1.383269)(6.1977777,1.576731)
			\pscircle[linecolor=black, linewidth=0.02, fillstyle=solid,fillcolor=black, dimen=outer](3.2155557,-1.4004912){0.065}
			\psline[linecolor=black, linewidth=0.04, arrowsize=0.05291667cm 2.0,arrowlength=1.4,arrowinset=0.0]{->}(0.02,-0.4143801)(0.02,0.7856199)
			\psline[linecolor=black, linewidth=0.04, arrowsize=0.05291667cm 2.0,arrowlength=1.4,arrowinset=0.0]{->}(0.02,-0.4143801)(1.22,-0.4143801)
			\psline[linecolor=black, linewidth=0.04](3.2155557,-1.4143801)(0.21555555,1.5856199)
			\psline[linecolor=black, linewidth=0.04](3.22,-1.4143801)(6.22,1.5856199)
			\rput[bl](3.4516666,-1.5238246){\normalsize{$x$}}
			\rput[bl](5.687778,0.39450878){\normalsize{propagation of information}}
			\pscircle[linecolor=black, linewidth=0.02, fillstyle=solid,fillcolor=black, dimen=outer](3.06,-0.8760468){0.065}
			\rput[bl](4.145,-1.0616024){\normalsize{$z$}}
			\pspolygon[linecolor=colour1, linewidth=0.02, fillstyle=solid,fillcolor=colour1](1.2822223,1.5678421)(3.2066667,-0.32549122)(5.113333,1.5678421)
			\psline[linecolor=black, linewidth=0.02](2.6688888,-0.8854912)(5.14,1.5722866)
			\psline[linecolor=black, linewidth=0.02](3.7444444,-0.8810468)(1.2733333,1.576731)
			\rput[bl](0.92,-0.2393801){\normalsize{$\vec{y}$}}
			\rput[bl](0.17,0.5656199){\normalsize{$y^0$}}
			\rput[bl](2.7744443,0.9345088){\normalsize{$\mathcal{I}^\vee(\tilde{x})$}}
			\psbezier[linecolor=black, linewidth=0.02, arrowsize=0.05291667cm 2.0,arrowlength=1.4,arrowinset=0.0]{->}(4.077778,-1.0010468)(3.8777778,-1.1121578)(3.5711112,-1.2543801)(3.1933334,-0.947713436550564)
			\psline[linecolor=black, linewidth=0.04, arrowsize=0.05291667cm 2.0,arrowlength=1.4,arrowinset=0.0]{->}(4.188889,0.13228656)(4.9844446,0.94117546)
			\psline[linecolor=black, linewidth=0.04, arrowsize=0.05291667cm 2.0,arrowlength=1.4,arrowinset=0.0]{->}(2.1088889,0.14117545)(1.3133334,0.95006436)
			\psbezier[linecolor=black, linewidth=0.02, arrowsize=0.05291667cm 2.0,arrowlength=1.4,arrowinset=0.0]{->}(5.6244445,0.47895324)(5.4244447,0.42339766)(5.117778,0.35228658)(4.74,0.5056198967827709)
			\pscircle[linecolor=black, linewidth=0.02, fillstyle=solid,fillcolor=black, dimen=outer](3.22,-0.338269){0.065}
			\rput[bl](3.433889,-0.43493566){\normalsize{$\tilde{x}$}}
			\end{pspicture}
		}
		\caption{The approximate center and the loss of access to information.}
		\label{figL2}
	\end{figure}%
	\item[{\rm{(ii)}}] The above discussion can be summarized formally in a more mathematical language as
\[ 
	\begin{split}
&F^{\varepsilon}(z)\in \big(\mathcal{W}^{\varepsilon,\vee}(\tilde{x})\big)'=\big(\mathcal{E}_M^{\varepsilon,\vee}(\tilde{x})\big)'\\[0.2em]
&\mbox{(approximately as $\varepsilon\searrow 0$) for any }z\in \mathcal{I}^\vee(x)\cap \mathcal{I}^\wedge(\tilde{x})\:,
\end{split}
\]
where by ``approximately as $\varepsilon \searrow 0$'' we mean that we disregard all
terms involving positive powers of~$\varepsilon$ (as on the right side of~\eqref{Fiota}).
Indeed, keeping in mind that 
	$$
	F^\varepsilon(z)\in \overline{\mathcal{W}^{\varepsilon,\vee}(\tilde{x})}= \mathcal{E}_M^{\varepsilon,\vee}(x)\quad\mbox{for any }z\in \mathcal{I}^\vee(x)\cap \mathcal{I}^\wedge(\tilde{x}) \:,
	$$
	this implies that
	\begin{equation*}
	\begin{split}
	&F^\varepsilon(z)\in \mathcal{E}_M^{\varepsilon,\vee}(x)\cap \big(\mathcal{E}_M^{\varepsilon,\vee}(\tilde{x})\big)'\\[0.2em]
	&\mbox{(approximately as $\varepsilon\searrow 0$) for any }z\in \mathcal{I}^\vee(x)\cap \mathcal{I}^\wedge(\tilde{x}) \:.
	\end{split}
	\end{equation*}
The commutators of the elements of $\{F^\varepsilon(z)\:|\: z\in \mathcal{I}^\vee(x)\cap \mathcal{I}^\wedge(\tilde{x})\}$ are expected to involve contributions which do not vanish in the limit~$\varepsilon \searrow 0$;
an explicit example is the commutator~$[F^\varepsilon(z), F^\varepsilon(\tilde{z})]$ if~$z$ and~$\tilde{z}$
are lightlike separated. 
Even though we  are focussing on \textit{spacetime restricted} future algebras, the above properties show
some similarities with the second statement in the PDP~\eqref{PDP2}. \\[-0.3em]
	\item[{\rm{(iii)}}] Consider again two points $x$ and $\tilde{x}$ in Minkowski space as in point (i) or (ii). By construction, it follows that the corresponding future algebras are included one into the other
	$
	{\mathcal{W}}^{\varepsilon,\vee}(\tilde{x})\subseteq {\mathcal{W}}^{\varepsilon,\vee}(x).
	$
	This inclusion is \textit{proper},
	$$
	{\mathcal{W}}^{\varepsilon,\vee}(\tilde{x})\subsetneq {\mathcal{W}}^{\varepsilon,\vee}(x) \:.
	$$ 
This is a consequence of Theorem \ref{theoremexpvalue}.  Indeed, using the incompatibility
of the estimates in the two statements of Theorem~\ref{theoremexpvalue}, an
operator~$A_f^\varepsilon\in \mathcal{W}^{\varepsilon,\vee}(x)$ with~$f$ supported in a sufficiently small neighborhood which intersects the future null cone centered at~$\tilde{x}$ (as in the second statement of Theorem \ref{theoremexpvalue}) cannot belong to $\mathcal{W}^{\varepsilon,\vee}(\tilde{x})$.
	
	It is worth mentioning that the proper inclusion just discussed remains true in the limit $\varepsilon\searrow 0$
	when the regularization is removed.
	In order to explain this in more detail, we remark that the algebra $\A_\Omega^\varepsilon$ defined in Definition \ref{algebramink} can be proven to converge element-wise to well-defined {\em{unregularized'' algebras}}~$\A_\Omega$ in the limit~$\varepsilon\searrow 0$ (this is explained in detail in \cite[Section 5]{neumann}).
	Introducing the unregularized future algebra 
	$$
	\mathcal{W}^\vee(x):=\A_{\mathcal{I}^\vee(x)} \:,
	$$
 it can then be shown that, similar as in the regularized case above,
	$$
	{\mathcal{W}}^{\vee}(\tilde{x})\subsetneq {\mathcal{W}}^{\vee}(x) \:.
	$$
	In conclusion, we see that the spacetime restricted future algebras of time-like separated points are properly included one into the other. This is true also in the limit $\varepsilon\searrow 0$, in agreement with the first statement in the PDP~\eqref{PDP2} (again, we remark that here we are focussing on the \textit{spacetime restricted} future algebras).
	\\[-0.3em]
	
	\item[{\rm{(iv)}}] The estimate of the commutator in Theorem \ref{prpiotaes} also holds in the particular case
	when~$\Omega$ is chosen as the open future light cone~$\mathcal{I}^\vee(x)$. To some extent,
	the discussion in point (i) also applies in the degenerate case $\tilde{x}=x$. Since
	$$
	F^\varepsilon(x)\in \overline{\mathcal{W}^{\varepsilon,\vee}(x)}=\mathcal{E}_M^{\varepsilon,\vee}(x)
	$$
and
\[ 
	F^\varepsilon(x)\in \big(\mathcal{W}^{\varepsilon,\vee}(x)\big)'=\big(\mathcal{E}_M^{\varepsilon,\vee}(x)\big)' \quad\mbox{(approximately as $\varepsilon\searrow 0$)} \:,
\]
	we conclude that the point $F^\varepsilon(x)$ belongs to the \textit{center} of its own spacetime restricted future algebra
	$$
	F^\varepsilon(x)\in \big(\mathcal{E}_M^{\varepsilon,\vee}(x)\big)'\cap \mathcal{E}_M^{\varepsilon,\vee}(x)\quad\mbox{(approximately  as $\varepsilon\searrow 0$)} \:.
	$$\\[-0.8em]
	This can be interpreted as follows:
	\begin{center}
	\textit{The operator $F^\varepsilon(x)$ belongs to the center of \\ the 
	spacetime restricted future algebra.}
	\end{center}
In this sense, the operator $F^\varepsilon(x)$, which can be interpreted as a potential event in its own future, share some features with the notion of an \textit{event that happens}, as introduced in Definition~\ref{ETHevent}. 
\end{itemize}
\vspace{0.1cm}
\noindent
We finally remark that, when interpreting the above results,
one should always keep in mind that the algebras in Minkowski space
describe the {\em{non-interacting Minkowski vacuum}}.
Therefore, all phenomena for which an interaction is essential
(like Born's rule or the ``collapse'' of the wave function in a measurement process
or when ETH events happen) cannot be analyzed in these examples.

\section{Outlook}\label{sec:outlook}
In this section we shall discuss the prospects of CFS theory for interacting quantum systems.
Since this is the subject of ongoing research (for some first steps, see~\cite{perturb, fockbosonic}),
we can only provide a tentative discussion of these matters, trying to indicate where further research might lead one to.
Thus, our discussion is necessarily speculative.

Before beginning to compare the material in Section 5 with the ETH approach, we
point out that the research initiated in~\cite{fockbosonic} might indicate that CFS theory
will give rise to an (interacting) Quantum Field Theory if the causal action principle
is analyzed for general nonlinear perturbations of
 measures describing Minkowski space. In particular, in~\cite{fockbosonic} it is shown that,
in the so-called {\em{holomorphic approximation}}, the dynamics predicted by such measures 
can be described by a unitary time evolution on Fock space.
One of the tasks is thus to justify the holomorphic approximation.
The strategy is to use a procedure called {\em{microscopic mixing}} which was first introduced in~\cite{qft}.
This procedure starts with $N$ copies of Minkowski space with different field configurations. 
Each of them is encoded in a measure satisfying the Euler-Lagrange equations of a causal fermion system, as 
discussed in Section~\ref{sec:CFS}. One then takes a convex combination of these $N$ measures.
In~\cite[Section~1.5.3]{cfs} it is argued that this procedure decreases the causal action.
Therefore, the resulting measure is a good candidate for an approximate minimizer.
By perturbing this measure (as is done systematically in~\cite[Sections~5 and~8]{perturb}),
one can decrease the causal action further to obtain again a solution of the Euler-Lagrange equations.
Clearly, the resulting measure has a very complicated structure.
The goal is to describe it effectively by a density operator on the tensor product of a bosonic and
fermionic Fock space. Further details need to be worked out.

To conclude, we wish to draw the readers' attention to two aspects regarding the connection between 
CFS theory and the ETH approach to QT worth thinking about.

The first aspect concerns the description of {\em{time evolution}}. In the ETH approach,
the (Heisenberg picture) \textit{dynamics of operators} is deterministic; (it is given by a semi-group of 
$^*$-endomorphisms from algebras $\mathcal{E}_{\geq t}$ to subalgebras $\mathcal{E}_{\geq t'}$, 
for $t'>t$). The \textit{dynamics of states} is, however, 
fundamentally \textit{probabilistic};  it is deterministic only as long as no event happens,
whereas if an event happens the change in the state of the system can only be predicted probabilistically. 
In contrast, as already mentioned in Section~\ref{secdescfs},
it has been proven in~\cite{linhyp} that, in CFS theory, the dynamics of linearized perturbations of measures describing
the Minkowski vacuum is well-posed and hence \textit{deterministic}, provided that the hyperbolicity conditions hold.
As discussed in Section~\ref{secdesdiscuss}, this major difference might disappear
if the nonlinear dynamics of causal fermion systems is analyzed.
Indeed, it is conceivable that the nonlinear time evolution described by the causal action principle
might ``trigger a collapse'' of the state vector leading to
a probabilistic time evolution, just as in the ETH approach. A preliminary proposal in this direction is made in~\cite[Chapter~5]{kleinerdr}.
Following the procedure in~\cite{fockbosonic}, such nonlinear evolutions can only be described within 
the many-particle formalism of Quantum Field Theory.
This is in accordance with the ETH approach, where it is argued that a consistent extension of quantum theory apt to describe measurements is necessarily a quantum theory of systems with infinitely many degrees of freedom exhibiting interactions and many-particle effects. 

The second aspect is related to the {\em{dichotomy between past and future}} in the ETH approach and to how one might think of events in CFS theory. The clear distinction between past and future is a characteristic feature of the ETH approach. The past consists of all events that actually happened, while the future should be thought of as the ensemble of all potential events that might happen. Clearly not everything that \textit{might} happen in the future \textit{will} happen, i.e., while the past is factual, the future is probabilistic.
This raises the question of how one might be able to account for these features in CFS theory, and of what might have to be modified to accommodate them? 
In~\cite{fockbosonic} it is shown that non-linear perturbations can be associated to a
state in a Fock space. The ETH approach teaches us that the state of a system changes nonlinearly when 
an event occurs. This suggests that one may want to investigate the following modification of CFS theory: 
One may attempt to split the universal measure $\rho$ into two pieces
\[ \rho=\rho_{\text{past}}+\rho_{\text{future}} \:, \]
where $\rho_{\text{past}}$ represents the unique history of the system and $\rho_{\text{future}}$ describes a mixture 
of all possible futures. The minimization in the causal action principle is then carried out for variations of 
$\rho_{\text{future}}$, with $\rho_{\text{past}}$ \textit{fixed}, and the minimization is over all measures 
$\rho_{\text{future}}$ in the future of $\rho_{\text{past}}$ in $\F$, 
defined as follows
(similar constrained variational principles have been studied mathematically in~\cite{cauchy}).
\begin{Def} {\bf{(Future of $\rho_{\text{past}}$)}}
\[ \F^\vee(\rho_{\text{\rm{past}}}):= \left\{x\in \F \left|\begin{array}{l}xy \text{ spacelike or }\\
    xy \text{ causal and }\mathcal{C}(x,y)>0 \end{array}\,\, \forall y \in \supp(\rho_{\text{\rm{past}}})\right.\right\} \]
\end{Def} \noindent
When an event occurs, a piece of one of the possible futures in the microscopic mixing
corresponding to $\rho_{\text{future}}$ is 
added to $\rho_{\text{past}}$, and the action principle then determines a new $\rho_{\text{future}}$. This gives 
a change in the state as required when an event occurs in the ETH approach to QT. The history of the universe, 
i.e., spacetime with all the structures therein, is built, event by event, from a probabilistic future.
It is unclear, at present, whether in such a scenario one should think of events as being represented by 
individual spacetime point operators or by collections thereof. This is partially due to the fact that we switch 
back and forth between a fundamental description, on the side of causal fermion systems, and effective descriptions. 
It will require a substantial amount of further research to put these ideas on firm grounds. 

\section{Conclusion}\label{sec:conclusion}
In the present paper we have given an overview of the state of research regarding the compatibility of CFS theory with the ETH approach to QT. 

We have presented an introduction to both frameworks and their mathematical structures, explaining that each of them addresses one conceptual problem in fundamental theoretical physics. It turns out that there are substantial similarities between the two theories. However, there are also some important, although quite subtle differences. We have presented the structural similarities and differences in a detailed overview in order to lay a 
basis for the subsequent discussion of recent mathematical results. The survey of these recent mathematical results shows that the relationship between the two theories is quite subtle even in the most simple example of a causal 
fermion system describing  the Minkowski vacuum. 
Given the current status of research, a definitive answer to the question whether the two 
theories can be merged to a complete and consistent theory unifying QT with a theory of Gravitation can obviously not be given, yet.

In the last section we have discussed a number of suggestions regarding possible future strategies to reach this goal. 
These suggestions are, however, rather speculative and should be taken with a grain of skepticism.

\Thanks {{\em{Acknowledgments:}}} 
C.F.P.\ is funded by the SNSF grant  P2SKP2 178198. 
We are grateful for support by the Vielberth Foundation, Regensburg, and ETH Z\"urich. We would like to thank the organizers of the
conference ``Progress and Visions in Quantum Theory in View of Gravity''
held in October 2018 at the Max Planck Institute for Mathematics in the Sciences, Leipzig,
for stimulating fruitful scientific interchange.
Finally, we are grateful to the referee for helpful comments on the manuscript.


\begin{thebibliography}{10}

\bibitem{cfsweblink}
\emph{Link to web platform on causal fermion systems:
  www.causal-fermion-system.com}.

\bibitem{sphere}
L.~B\"auml, F.~Finster, H.~von~der Mosel, and D.~Schiefeneder, \emph{Singular
  support of minimizers of the causal variational principle on the sphere},
  arXiv:1808.09754 [math.CA], Calc. Var. Partial Differential Equations
  \textbf{58} (2019), no.~6, 205.

\bibitem{blanchard2016garden}
P.~Blanchard, J.~Fr{\"o}hlich, and B.~Schubnel, \emph{A “garden of forking
  paths” -- the quantum mechanics of histories of events}, Nuclear Physics B
  \textbf{912} (2016), 463--484.

\bibitem{buchholz2014new}
D.~Buchholz and J.E. Roberts, \emph{New light on infrared problems: {S}ectors,
  statistics, symmetries and spectrum}, Commun. Math. Phys. \textbf{330}
  (2014), no.~3, 935--972.

\bibitem{BMW}
L.J. Bunce and J.D.~Maitland Wright, \emph{The {M}ackey-{G}leason problem},
  Bull. Amer. Math. Soc. \textbf{26} (1992), 288--293.

\bibitem{jacobson}
E.~Curiel, F.~Finster, and J.M. Isidro, \emph{Two-dimensional area and matter
  flux in the theory of causal fermion systems}, arXiv:1910.06161 [math-ph], to
  appear in Internat. J. Modern Phys. D (2020).

\bibitem{linhyp}
C.~Dappiaggi and F.~Finster, \emph{Linearized fields for causal variational
  principles: {E}xistence theory and causal structure}, arXiv:1811.10587
  [math-ph], Methods Appl. Anal. \textbf{27} (2020), no.~1, 1--56.

\bibitem{doplicher}
S.~Doplicher, K.~Fredenhagen, and J.E. Roberts, \emph{The quantum structure of
  spacetime at the {P}lanck scale and quantum fields}, Commun. Math. Phys.
  \textbf{172} (1995), no.~1, 187--220.

\bibitem{dvurevcenskij}
A.~Dvure\v{c}enskij, \emph{Gleason's {T}heorem and its {A}pplications},
  Mathematics and its Applications (East European Series), vol.~60, Kluwer
  Academic Publishers Group, Dordrecht; Ister Science Press, Bratislava, 1993.

\bibitem{pfp}
F.~Finster, \emph{The {P}rinciple of the {F}ermionic {P}rojector},
  hep-th/0001048, hep-th/0202059, hep-th/0210121, AMS/IP Studies in Advanced
  Mathematics, vol.~35, American Mathematical Society, Providence, RI, 2006.

\bibitem{reg}
\bysame, \emph{On the regularized fermionic projector of the vacuum},
  arXiv:math-ph/0612003, J. Math. Phys. \textbf{49} (2008), no.~3, 032304, 60.

\bibitem{continuum}
\bysame, \emph{Causal variational principles on measure spaces},
  arXiv:0811.2666 [math-ph], J. Reine Angew. Math. \textbf{646} (2010),
  141--194.

\bibitem{qft}
\bysame, \emph{Perturbative quantum field theory in the framework of the
  fermionic projector}, arXiv:1310.4121 [math-ph], J. Math. Phys. \textbf{55}
  (2014), no.~4, 042301.

\bibitem{cfsrev}
\bysame, \emph{Causal fermion systems -- an overview}, arXiv:1505.05075
  [math-ph], {Q}uantum {M}athematical {P}hysics: A {B}ridge between
  {M}athematics and {P}hysics (F.~Finster, J.~Kleiner, C.~Röken, and
  J.~Tolksdorf, eds.), Birkh\"auser Verlag, Basel, 2016, pp.~313--380.

\bibitem{cfs}
\bysame, \emph{The {C}ontinuum {L}imit of {C}ausal {F}ermion {S}ystems},
  arXiv:1605.04742 [math-ph], Fundamental Theories of Physics, vol. 186,
  Springer, 2016.

\bibitem{dice2018}
\bysame, \emph{Causal fermion systems: {D}iscrete space-times, causation and
  finite propagation speed}, arXiv:1812.00238 [math-ph], J. Phys.: Conf. Ser.
  \textbf{1275} (2019), 012009.

\bibitem{perturb}
\bysame, \emph{Perturbation theory for critical points of causal variational
  principles}, arXiv:1703.05059 [math-ph], Adv. Theor. Math. Phys. \textbf{24}
  (2020), no.~3, 563--619.

\bibitem{lqg}
F.~Finster and A.~Grotz, \emph{A {L}orentzian quantum geometry},
  arXiv:1107.2026 [math-ph], Adv. Theor. Math. Phys. \textbf{16} (2012), no.~4,
  1197--1290.

\bibitem{cauchy}
\bysame, \emph{On the initial value problem for causal variational principles},
  arXiv:1303.2964 [math-ph], J. Reine Angew. Math. \textbf{725} (2017),
  115--141.

\bibitem{rrev}
F.~Finster, A.~Grotz, and D.~Schiefeneder, \emph{Causal fermion systems: A
  quantum space-time emerging from an action principle}, arXiv:1102.2585
  [math-ph], {Q}uantum {F}ield {T}heory and {G}ravity (F.~Finster, O.~M\"uller,
  M.~Nardmann, J.~Tolksdorf, and E.~Zeidler, eds.), Birkh\"auser Verlag, Basel,
  2012, pp.~157--182.

\bibitem{review}
F.~Finster and M.~Jokel, \emph{Causal fermion systems: An elementary
  introduction to physical ideas and mathematical concepts}, arXiv:1908.08451
  [math-ph], {P}rogress and {V}isions in {Q}uantum {T}heory in {V}iew of
  {G}ravity (F.~Finster, D.~Giulini, J.~Kleiner, and J.~Tolksdorf, eds.),
  Birkh\"auser Verlag, Basel, 2020, pp.~63--92.

\bibitem{fockbosonic}
F.~Finster and N.~Kamran, \emph{Complex structures on jet spaces and bosonic
  {F}ock space dynamics for causal variational principles}, arXiv:1808.03177
  [math-ph], to appear in Pure Appl. Math. Q. (2020).

\bibitem{dice2014}
F.~Finster and J.~Kleiner, \emph{Causal fermion systems as a candidate for a
  unified physical theory}, arXiv:1502.03587 [math-ph], J. Phys.: Conf. Ser.
  \textbf{626} (2015), 012020.

\bibitem{jet}
\bysame, \emph{A {H}amiltonian formulation of causal variational principles},
  arXiv:1612.07192 [math-ph], Calc. Var. Partial Differential Equations
  \textbf{56:73} (2017), no.~3, 33.

\bibitem{noncompact}
F.~Finster and C.~Langer, \emph{Causal variational principles in the
  $\sigma$-locally compact setting: {E}xistence of minimizers},
  arXiv:2002.04412 [math-ph], to appear in Adv. Calc. Var. (2020).

\bibitem{neumann}
F.~Finster and M.~Oppio, \emph{Local algebras for causal fermion systems in
  {M}inkowski space}, arXiv:2004.00419 [math-ph] (2020).

\bibitem{support}
F.~Finster and D.~Schiefeneder, \emph{On the support of minimizers of causal
  variational principles}, arXiv:1012.1589 [math-ph], Arch. Ration. Mech. Anal.
  \textbf{210} (2013), no.~2, 321--364.

\bibitem{frohlich2016quest}
J.~Fr{\"o}hlich, \emph{The quest for laws and structure}, Mathematics and
  Society (2016), 101--129.

\bibitem{froehlich2019review}
\bysame, \emph{A brief review of the ``{ETH}-approach to quantum mechanics''},
  arXiv:1905.06603 [quant-ph] (2019).

\bibitem{froehlich2019relativistic}
\bysame, \emph{Relativistic quantum theory}, arXiv:1912.00726 [quant-ph]
  (2019).

\bibitem{F}
\bysame, \emph{``{D}iminishing potentialities'', entanglement, ``purification''
  and the emergence of \underline{events} in quantum mechanics -- a simple
  model}, Sect. 5.6 of Notes for a course on Quantum Theory at LMU-Munich
  (Nov./Dec. 2019).

\bibitem{frohlich2015math}
J.~Fr{\"o}hlich and B.~Schubnel, \emph{Quantum probability theory and the
  foundations of quantum mechanics}, arXiv:1310.1484 [quant-ph], The Message of
  Quantum Science, Springer, 2015, pp.~131--193.

\bibitem{Gleason}
A.M. Gleason, \emph{Measures on the closed subspaces of a {H}ilbert space}, J.
  Math. Mech. \textbf{6} (1957), 885--893.

\bibitem{hegerfeldt1974remark}
G.C Hegerfeldt, \emph{Remark on causality and particle localization}, Physical
  Review D \textbf{10} (1974), no.~10, 3320.

\bibitem{kleinerdr}
J.~Kleiner, \emph{{Dynamics of Causal Fermion Systems -- Field Equations and
  Correction Terms for a New Unified Physical Theory}}, Dissertation,
  Universit\"at Regensburg, 2017.

\bibitem{lin-huaxin}
H.~Lin, \emph{Almost commuting selfadjoint matrices and applications}, Operator
  algebras and their applications ({W}aterloo, {ON}, 1994/1995), Fields Inst.
  Commun., vol.~13, Amer. Math. Soc., Providence, RI, 1997, pp.~193--233.

\bibitem{thaller}
B.~Thaller, \emph{The {D}irac {E}quation}, Texts and Monographs in Physics,
  Springer-Verlag, Berlin, 1992.

\end{thebibliography}
\providecommand{\bysame}{\leavevmode\hbox to3em{\hrulefill}\thinspace}
\providecommand{\MR}{\relax\ifhmode\unskip\space\fi MR }
\providecommand{\MRhref}[2]{%
  \href{http://www.ams.org/mathscinet-getitem?mr=#1}{#2}
}
\providecommand{\href}[2]{#2}

\end{document}